\documentclass[a4paper,fleqn]{cas-dc}

\usepackage{amsmath,amssymb,amsfonts}
\usepackage{graphicx, booktabs, xcolor, float}
\usepackage{caption, subcaption}
\usepackage{bm}       
\usepackage[capitalise]{cleveref} 
\usepackage{hyperref} 
\usepackage{placeins}

\usepackage{tikz}
\usetikzlibrary{positioning, arrows.meta, calc, shapes.geometric, fit, backgrounds, shadows}
\definecolor{net_input}{RGB}{0, 150, 136}
\definecolor{net_hidden}{RGB}{96, 125, 139}
\definecolor{net_output}{RGB}{63, 81, 181}
\definecolor{physics_bg}{RGB}{227, 242, 253}
\definecolor{um_bg}{RGB}{243, 229, 245}
\definecolor{um_border}{RGB}{156, 39, 176}
\definecolor{process_box}{RGB}{255, 255, 255}

\providecommand{\pd}[2]{\frac{\partial #1}{\partial #2}}
\providecommand{\dd}[2]{\frac{d #1}{d #2}}
\providecommand{\vect}[1]{\mathbf{#1}}
\providecommand{\Uvec}{\vect{U}}
\providecommand{\Fvec}{\vect{F}}
\providecommand{\Gvec}{\vect{G}}
\ifx\remark\undefined
  \newtheorem{remark}{Remark}
\fi

\begin{document}

\let\WriteBookmarks\relax
\def\floatpagepagefraction{1}
\def\textpagefraction{.001}
\shorttitle{Spatio-Temporal Uncertainty-Modulated PINNs}
\shortauthors{D. Zhao et al.}


\title [mode = title]{Spatio-Temporal Uncertainty-Modulated Physics-Informed Neural Networks for Solving Hyperbolic Conservation Laws with Strong Shocks}

\author[1]{Darui Zhao}[style=chinese, orcid=0009-0003-6352-6317]
\fnmark[1] 

\author[1]{Ze Tao}[style=chinese, orcid=0009-0004-0202-3641]
\fnmark[1] 

\author[1]{Fujun Liu}[style=chinese, orcid=0000-0002-8573-450X]
\cormark[1] 
\ead{fjliu@cust.edu.cn} 

\address[1]{Nanophotonics and Biophotonics Key Laboratory of Jilin Province, School of Physics, Changchun University of Science and Technology, Changchun 130022, P.R. China}

\cortext[cor1]{Corresponding author.}
\fntext[fn1]{These authors contributed equally to this work.}
\begin{highlights}
    \item Spatio-Temporal UM-PINN resolves gradient pathology in hyperbolic systems.
    \item Dual modulation via spatial masking and uncertainty weighting balances loss. 
    \item Accurately captures post-shock oscillations in Shu-Osher and complex shock-interaction structures in the 2D Riemann problem. 
    \item Superior robustness and accuracy compared to LRA and GradNorm methods.
\end{highlights}
\begin{graphicalabstract}
\includegraphics[width=\textwidth]{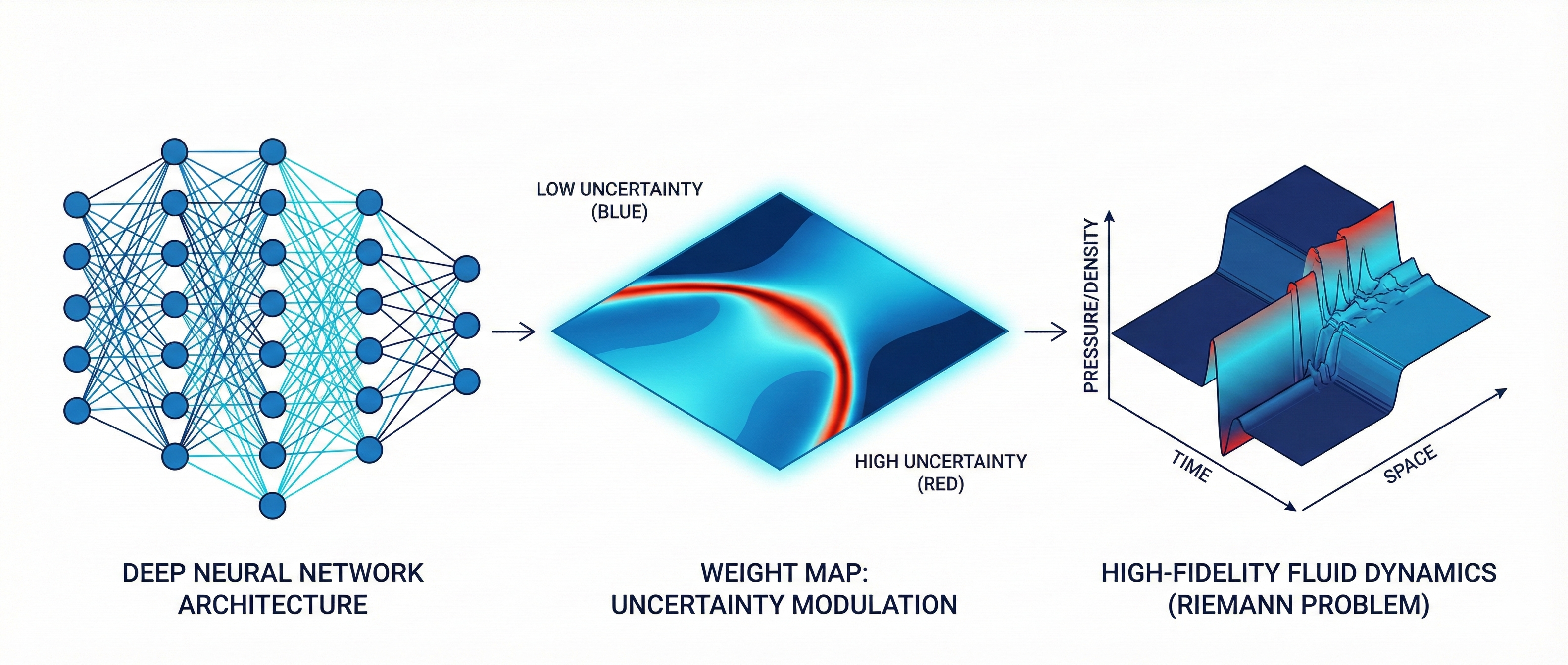}
\end{graphicalabstract}

\begin{abstract}
Physics-Informed Neural Networks (PINNs) often struggle to resolve shock-dominated hyperbolic conservation laws because the optimization is dominated by highly localized and highly unbalanced gradients near discontinuities. In this work, we study a Spatio-Temporal Uncertainty-Modulated PINN (UM-PINN) that combines two complementary mechanisms: a gradient-based spatial modulation term that attenuates extreme local residual spikes, and a homoscedastic uncertainty-based task modulation term that adaptively balances PDE, initial-condition, and boundary-condition losses. The resulting formulation remains a global-in-time coordinate-based PINN, while using Sobol-sequence sampling to improve the coverage of collocation points in the spatio-temporal domain. We evaluate the method on the 1D Sod shock tube, the 1D Shu-Osher problem, and a 2D Riemann problem, and we additionally conduct parameter-sensitivity, Sobol-versus-random, component-ablation, causal-baseline, and SOTA-oriented comparison studies. Across these tests, UM-PINN consistently improves training stability relative to standard PINN, LRA, and GradNorm baselines, achieves the best overall performance on the Shu-Osher benchmark, and delivers leading or near-leading accuracy on Sod among the additional shock-oriented comparators. These results demonstrate that UM-PINN provides an effective local-global dual-modulation strategy for shock-dominated PINN training.
\end{abstract}


\maketitle

\section{Introduction}
\label{sec:intro}

Hyperbolic conservation laws, particularly the Euler equations, serve as the foundational mathematical models for simulating supersonic aerodynamics, explosion mechanics, and astrophysical phenomena.
While accurately resolving shock waves and contact discontinuities is crucial for applications ranging from aircraft design to large-scale cosmic evolution simulations, capturing these discontinuous solutions remains a central challenge in computational fluid dynamics (CFD).
These physical phenomena are characterized by extremely high spatial gradients and strong nonlinear coupling.
Consequently, traditional mesh-based CFD pipelines can become computationally demanding when mesh generation is difficult or when the governing flow problem becomes high-dimensional \cite{Cai2021}.
In recent years, Physics-Informed Neural Networks (PINNs) \cite{Raissi2019PINN,Tao_2026,lu2021deeponet,jagtap2020xpinns} have emerged as a promising mesh-free paradigm driven by the rise of deep learning.
Researchers have leveraged the nonlinear fitting capabilities of neural networks to embed physical equations directly into loss functions via automatic differentiation (AD).
This approach has yielded significant progress in solving both forward simulations and inverse parameter inversion problems \cite{Cai2021, Mao2020, Raissi2020, Karniadakis2021}.
Despite these advancements, existing PINN frameworks still face limitations when dealing with strong shock waves.
A primary obstacle is the phenomenon known as "gradient pathology," where a severe magnitude imbalance exists between the partial differential equation (PDE) residual term and the initial condition (IC) term.
As a result, standard optimization algorithms often become trapped in local optima, causing excessive smoothing of shock profiles or non-physical Gibbs oscillations \cite{Fuks2020, DeRyck2022, Krishnapriyan2021, Jagtap2020}.
Such limitations challenge the reliability of PINNs for industrial-grade hypersonic flow simulations.
To improve robustness, our research incorporates homoscedastic aleatoric uncertainty to dynamically adjust the loss weights across different tasks. The resulting task-modulation mechanism treats the loss weights as learnable noise variances rather than fixed hyperparameters, which helps relieve the scale conflict between PDE, initial-condition, and boundary-condition constraints \cite{McClenny2023, Kendall2018, Psaros2022, Jin2021}.

On top of this global loss balancing, we introduce a gradient-based spatial modulation term to attenuate extreme residual spikes near shocks. The combination of local residual modulation and global task modulation forms a local-global dual modulation mechanism for shock-dominated PINN training. Rather than presenting these components as unrelated add-ons, we study how their interaction affects optimization stability, solution sharpness, and robustness across 1D and 2D Euler benchmarks.

To further strengthen the methodological and empirical validation of the proposed method, this study additionally reports (i) sensitivity analyses for the spatial-modulation parameters $\alpha$ and $\beta$, (ii) a Sobol-versus-random sampling ablation, (iii) a component ablation for the two modulation mechanisms, (iv) a comparison against a causal-loss baseline, and (v) SOTA-oriented baseline comparisons on the 1D shock problems. Together, these analyses establish the robustness, mechanism-level complementarity, and comparative effectiveness of UM-PINN for shock-dominated hyperbolic conservation laws.

The main contributions of this work are therefore summarized as follows: (1) a spatial residual modulation strategy that alleviates localized extreme gradients around discontinuities; (2) an uncertainty-based task modulation strategy that adaptively coordinates the competing PDE, IC, and BC objectives; (3) a unified UM-PINN framework that combines these two mechanisms as a dual modulation strategy for shock-dominated PINN training; and (4) an expanded numerical assessment showing that the method consistently improves training robustness, achieves the best overall performance on the Shu-Osher benchmark, and attains leading or near-leading accuracy on the Sod benchmark in the strengthened comparative evaluation.

\section{Mathematical Framework and Methodology}
\label{sec:math_framework}

We consider the two-dimensional (2D) Euler equations describing the motion of an inviscid, compressible fluid.
In conservative form, the system represents the conservation of mass, momentum, and energy:
\begin{equation}
	\frac{\partial \mathbf{U}}{\partial t} + \frac{\partial \mathbf{F}(\mathbf{U})}{\partial x} + \frac{\partial \mathbf{G}(\mathbf{U})}{\partial y} = 0,
	\label{eq:euler_2d_vec}
\end{equation}
where $\mathbf{U}$ is the vector of conserved variables, and $\mathbf{F}(\mathbf{U})$ and $\mathbf{G}(\mathbf{U})$ are the flux vectors in the $x$- and $y$-directions, respectively.
Their mathematical expressions are defined as follows:
\begin{equation}
	\mathbf{U} = \begin{bmatrix} \rho \\ \rho u \\ \rho v \\ E \end{bmatrix}, \quad
	\mathbf{F}(\mathbf{U}) = \begin{bmatrix} \rho u \\ \rho u^2 + p \\ \rho uv \\ (E + p)u \end{bmatrix}, \quad
	\mathbf{G}(\mathbf{U}) = \begin{bmatrix} \rho v \\ \rho uv \\ \rho v^2 + p \\ (E + p)v \end{bmatrix}.
	\label{eq:flux_vectors}
\end{equation}
where $\rho$ denotes the fluid density, $u$ and $v$ are the velocity components in the $x$- and $y$-directions, $p$ is the static pressure, and $E$ is the total energy per unit volume.
To close the system mathematically, an equation of state(EOS) is required to establish the relationship between pressure $p$ and the other conserved variables.
For an ideal gas, the thermodynamic relation follows:
\begin{equation}
	E = \frac{p}{\gamma - 1} + \frac{1}{2}\rho (u^2 + v^2).
\end{equation}
Consequently, the formula for pressure is given by:
\begin{equation}
	p = (\gamma - 1) \left( E - \frac{1}{2}\rho (u^2 + v^2) \right),
	\label{eq:eos}
\end{equation}
where $\gamma$ is the ratio of specific heats.
In the numerical experiments of this work, we set $\gamma = 1.4$ for the Sod and Shu-Osher problems, while for the 2D Riemann problem, the constant is selected according to the specific configuration.
For the one-dimensional cases involved in this study (e.g., Sod Shock Tube and Shu-Osher Problem), the physical fields exhibit translational invariance in the $y$-direction, implying $\partial/\partial y = 0$ and $v = 0$.
Under these conditions, Eq. \eqref{eq:euler_2d_vec} degenerates to the one-dimensional (1D) conservative form:
\begin{equation}
	\frac{\partial \mathbf{U}_{1D}}{\partial t} + \frac{\partial \mathbf{F}(\mathbf{U}_{1D})}{\partial x} = 0.
\end{equation}
The corresponding variables are simplified to:
\begin{equation}
	\mathbf{U}_{1D} = \begin{bmatrix} \rho \\ \rho u \\ E \end{bmatrix}, \quad
	\mathbf{F}(\mathbf{U}_{1D}) = \begin{bmatrix} \rho u \\ \rho u^2 + p \\ (E + p)u \end{bmatrix}.
\end{equation}

In the PINN framework, we employ a deep neural network $f_{\theta}$ to approximate the primitive variables.
This mapping can be represented as:
\begin{equation}
	\hat{\mathbf{y}}(t, x, y) = \begin{bmatrix} \hat{\rho} \\ \hat{u} \\ \hat{v} \\ \hat{p} \end{bmatrix} = f_{\theta}(t, x, y).
\end{equation}
Leveraging Automatic Differentiation (AD), we can precisely compute the derivatives of the network outputs with respect to the input coordinates.
The physical residuals $\mathcal{R}$ are defined as follows:
\begin{align}
	\mathcal{R}_{\rho} &:= \partial_t \hat{\rho} + \partial_x (\hat{\rho} \hat{u}) + \partial_y (\hat{\rho} \hat{v}) \nonumber \\
	\mathcal{R}_{\rho u} &:= \partial_t (\hat{\rho} \hat{u}) + \partial_x (\hat{\rho} \hat{u}^2 + \hat{p}) + \partial_y (\hat{\rho} \hat{u} \hat{v}) \nonumber \\
	\mathcal{R}_{\rho v} &:= \partial_t (\hat{\rho} \hat{v}) + \partial_x (\hat{\rho} \hat{u} \hat{v}) + \partial_y (\hat{\rho} \hat{v}^2 + \hat{p}) \label{eq:residuals} \\
	\mathcal{R}_{E} &:= \partial_t \hat{E} + \partial_x ((\hat{E} + \hat{p})\hat{u}) + \partial_y ((\hat{E} + \hat{p})\hat{v}) \nonumber
\end{align}
where $\hat{E}$ is computed from the predicted primitive variables using the EOS. To specifically handle the extreme gradient fluctuations near shock waves, we introduce a \textbf{Spatial Modulation} mechanism.
Inspired by the concept of flux limiters in traditional CFD, we define the modulated physical residuals $\hat{R}$ as:
\begin{equation}
    \hat{R} = \frac{1}{1 + \alpha \left| \nabla \hat{U} \right|^\beta} \odot R
\end{equation}
where $\left| \nabla \hat{U} \right|$ denotes the norm of the gradient of predicted conserved variables, and $\alpha, \beta$ are hyperparameters that control the degree of attenuation.
This mechanism prevents the optimization from being dominated by localized high-gradient spikes at discontinuities.
Mathematically, at shock locations where $|\nabla \hat{U}| \gg 1$, the modulation factor $\frac{1}{1+\alpha|\nabla\hat{U}|^\beta} \to 0$, effectively down-weighting the residual contribution from these extreme points.
This ensures a smoother loss landscape and prevents gradient explosion during backpropagation, analogous to how flux limiters suppress oscillations in traditional CFD.
Here, $\alpha$ controls the overall strength of the spatial modulation, while $\beta$ governs the nonlinear sensitivity of the attenuation factor to the local gradient magnitude. To clarify their roles, the sensitivity study in \Cref{subsec:alpha_beta} evaluates $\alpha \in \{0,0.5,1,2,5\}$ with $\beta=1.25$ fixed, and $\beta \in \{0.5,1,1.25,1.5,2\}$ with $\alpha=1$ fixed. The resulting trends show that removing the modulation weakens shock-resolving performance, whereas excessively strong modulation can also degrade accuracy.
These residual terms are incorporated into the loss function. When the neural network fully satisfies the governing equations, the residual vector $\mathbf{R} = [\mathcal{R}_{\rho}, \mathcal{R}_{\rho u}, \mathcal{R}_{\rho v}, \mathcal{R}_{E}]^T$ should converge to a zero vector.
Having established the governing equations and residual definitions, we now focus on the core components of the proposed UM-PINN framework.
This includes the Bayesian uncertainty weighting mechanism, the network architecture, the sampling strategy, and the optimization protocol.
In solving hyperbolic conservation laws characterized by shock waves, the gradients of the PDE residual terms at discontinuities are often orders of magnitude higher than those of the initial condition (IC) terms.
This severe imbalance often causes the optimizer to get stuck in local optima, a phenomenon associated with ``gradient pathology'' \cite{Wang2021Pathology}.
To address this, we formulate the training of PINNs as a multi-task learning problem under \textbf{homoscedastic aleatoric uncertainty}.
From a probabilistic perspective, we postulate that the prediction error for each task $i$ (e.g., PDE residuals, ICs, BCs) follows a Gaussian distribution with zero mean and variance $\sigma_i^2$.
The likelihood function can be expressed as:
\begin{equation}
	p(\mathbf{y} | f_\theta(\mathbf{x}), \sigma) = \mathcal{N}(f_\theta(\mathbf{x}), \sigma^2),
\end{equation}
where $f_\theta(\mathbf{x})$ denotes the network output and $\mathbf{y}$ represents the target values (typically zero for PDE residuals).
To simultaneously optimize multiple competing physical constraints, we seek to maximize the joint log-likelihood.
For independent tasks, minimizing the negative log-likelihood (NLL) yields the following objective:
\begin{equation}
	\mathcal{L}(\theta, \sigma) = \sum_{i} \left( \frac{1}{2\sigma_i^2} \mathcal{L}_i(\theta) + \log \sigma_i \right),
\end{equation}
where $\mathcal{L}_i(\theta)$ represents the original Mean Squared Error (MSE) for task $i$.
It is important to note that for the PDE task, $L_{PDE}(\theta)$ is computed based on the modulated residuals $\hat{R}$ rather than the raw residuals $R$.
Specifically:
\begin{equation}
    L_{PDE}(\theta) = \frac{1}{N_{PDE}} \sum_{j=1}^{N_{PDE}} \left\| \hat{R}(t_j, x_j, y_j) \right\|^2
\end{equation}
This dual-modulation strategy, combining \textbf{Spatial Modulation} for localized gradient balancing and \textbf{Task Modulation} via $s_i$ for global term weighting, constitutes the core of the UM-PINN framework.
To improve numerical stability and prevent potential division by zero, we introduce a learnable log-variance parameter $s_i := \log \sigma_i^2$.
Consequently, the total loss function for UM-PINN is defined as:
\begin{equation}
	\mathcal{L}_{total}(\theta, \mathbf{s}) = \sum_{i \in \mathcal{T}} \left( \frac{1}{2} e^{-s_i} \mathcal{L}_i(\theta) + \frac{1}{2} s_i \right),
	\label{eq:um_loss}
\end{equation}
where $\mathcal{T} = \{PDE, IC, BC\}$.
The rationale behind this design is twofold. First, the term $e^{-s_i}$ acts as an adaptive weight for automatic weight adaptation.
When the physical residual $\mathcal{L}_{PDE}$ is extremely large (e.g., near shocks), the network increases $s_i$ to reduce the weight of that term, preventing its gradients from overwhelming the optimization process.
Second, the term $\frac{1}{2} s_i$ serves as a regularization constraint, penalizing the network to prevent indefinite increases in uncertainty, thus avoiding the trivial solution where weights are set to zero to escape learning.
\begin{figure*}
\resizebox{\textwidth}{!}{   
\begin{tikzpicture}[
    node distance=1.2cm and 2.0cm,
    neuron/.style={circle, draw=white, line width=1pt, minimum size=0.85cm, inner sep=0pt, drop shadow, font=\small},
    input/.style={neuron, fill=net_input, text=white},
    hidden/.style={neuron, fill=net_hidden, text=white},
    output/.style={neuron, fill=net_output, text=white},
    connect/.style={-latex, gray!50, line width=0.5pt},
    block/.style={rectangle, draw=gray!30, fill=process_box, rounded corners=3pt, minimum height=1cm, align=center, drop shadow, font=\small},
    um_block/.style={rectangle, draw=um_border, fill=white, rounded corners=3pt, align=center, drop shadow, line width=0.8pt},
    title/.style={font=\bfseries\small, align=center}
]

    \node[input] (t) at (0, 1.2) {$t$};
    \node[input] (x) at (0, -0.2) {$x$};
    \node[left=0.4cm of t, font=\bfseries] {Input};

    \foreach \h in {1,2,3} {
        \foreach \i in {1,2,3,4} {
            \node[hidden] (h\h_\i) at (1.8*\h, 2.2 - 1.0*\i) {};
        }
    }
    
    \foreach \source in {t, x} \foreach \dest in {1,2,3,4} \draw[connect] (\source) -- (h1_\dest);
    \foreach \layer in {1,2} {
        \pgfmathtruncatemacro{\nextlayer}{\layer+1}
        \foreach \source in {1,2,3,4} \foreach \dest in {1,2,3,4} \draw[connect] (h\layer_\source) -- (h\nextlayer_\dest);
    }

    \node[output] (rho) at (7.2, 1.2) {$\rho$};
    \node[output] (u) at (7.2, 0.2) {$u$};
    \node[output] (p) at (7.2, -0.8) {$p$};
    \node[below=0.1cm of p, font=\bfseries\footnotesize, color=net_output] {Prediction $\mathcal{U}_\theta$};
    \foreach \source in {1,2,3,4} \foreach \dest in {rho, u, p} \draw[connect] (h3_\source) -- (\dest);
    \node[block, right=1.5cm of u, text width=2.2cm] (ad) {
        \textbf{Auto Diff} \\ 
        $\partial_t, \partial_x$
    };
    \node[block, below=0.8cm of ad, text width=2.2cm, fill=physics_bg] (pde) {
        \textbf{Euler PDE} \\
        Residuals $\mathcal{R}$
    };
    \draw[connect, line width=1pt] (rho) -| (ad.north);
    \draw[connect, line width=1pt] (u) -- (ad.west);
    \draw[connect, line width=1pt] (p) -| (ad.south);
    \draw[connect, line width=1pt] (ad) -- (pde);

    \node[right=2.0cm of ad] (um_center) {};
    \node[um_block, above=0.3cm of um_center, text width=3.5cm] (spatial) {
        \textbf{1. Spatial Modulation} \\
        (Gradient Masking) \\
        $\hat{\mathcal{R}} = \frac{1}{1+\alpha|\nabla\mathcal{U}|^\beta} \odot \mathcal{R}$
    };
    \node[um_block, below=0.8cm of spatial, text width=3.5cm] (task) {
        \textbf{2. Task Modulation} \\
        (Uncertainty Weighting) \\
        $\sum (e^{-s_i} \mathcal{L}_i + s_i)$
    };
    \begin{scope}[on background layer]
        \node[fit=(spatial)(task), fill=um_bg, draw=um_border, dashed, rounded corners, line width=1pt, inner sep=10pt] (um_box) {};
        \node[above, font=\bfseries, text=um_border] at (um_box.north) {Uncertainty-Modulated (UM) Module};
    \end{scope}

    \draw[connect] (ad.east) -- ++(0.5,0) |- (spatial.west) node[midway, above, yshift=2pt] {$\nabla \mathcal{U}$};
    \draw[connect] (pde.east) -| (spatial.south);

    \draw[connect, line width=1pt] (spatial.south) -- (task.north) node[midway, right, font=\tiny] {$\hat{\mathcal{L}}_{PDE}$};
    \node[below=1.5cm of pde, font=\footnotesize] (data) {Ground Truth $\mathcal{U}_0$};
    \draw[connect, dashed] (data) -| (task.south west) node[midway, below, font=\tiny] {IC Error};
    \node[circle, fill=black!80, text=white, right=0.8cm of um_box, font=\bfseries] (opt) {Opt};
    \draw[connect, line width=1.5pt] (um_box.east) -- (opt);
    \draw[-latex, dashed, line width=1pt, gray] (opt.south) |- (0, -3.5) node[pos=0.2, below] {Gradient Descent} -| (x.south);
\end{tikzpicture}
}
    \caption{Schematic architecture of the Uncertainty-Modulated Physics-Informed Neural Network (UM-PINN).
    The network predicts the primitive variables $(\rho, u, p)$ from spatio-temporal coordinates $(t, x)$.
    The core innovation is the \textbf{Uncertainty-Modulated (UM) Module}}
    \label{fig:um-pinn}
\end{figure*}

We employ a fully connected (FC) deep neural network to approximate the mapping $(t, x, y) \to (\rho, u, v, p)$.
An FC network consists of multiple layers where each neuron in layer $l$ is connected to every neuron in layer $l+1$.
For a hidden layer with input $\mathbf{h}^{(l-1)}$, the output is computed as $\mathbf{h}^{(l)} = \sigma(\mathbf{W}^{(l)} \mathbf{h}^{(l-1)} + \mathbf{b}^{(l)})$, where $\mathbf{W}^{(l)}$ is the weight matrix, $\mathbf{b}^{(l)}$ is the bias vector, and $\sigma$ is the activation function.
To ensure the predictions adhere to physical laws, specific constraints are applied at the output layer.
For instance, we apply the \texttt{Softplus} activation function to the predicted density $\hat{\rho}$ and pressure $\hat{p}$.
The Softplus function is defined as $\text{Softplus}(x) = \ln(1 + e^x)$, which is a smooth approximation to the Rectified Linear Unit (ReLU) function and guarantees strictly positive outputs, ensuring physical plausibility for thermodynamic quantities like density and pressure.
Furthermore, the hidden layers use the \texttt{Tanh} activation function, whose smooth differentiability is convenient for automatic-differentiation-based evaluation of the first-order Euler residuals.
To capture sharp transitions in the spatiotemporal domain more efficiently, we discard uniform random sampling in favor of Sobol Sequences \cite{Sobol1967}, a type of Quasi-Monte Carlo method.
The primary advantage of Sobol sequences is their low-discrepancy property.
Compared to pseudo-random sampling, Sobol points are distributed more uniformly across the domain, avoiding clustering.
This provides a more uniform coverage of the spatio-temporal domain, reducing local clustering and coverage gaps in the collocation set, which can improve the approximation quality of the residual loss and benefit training efficiency in shock-dominated problems.
In practice, this improvement should be interpreted as a coverage advantage rather than as a universal guarantee of better final accuracy. The Sobol-versus-random ablation in \Cref{subsec:sobol_random} shows that the gain is case-dependent, but that low-discrepancy coverage can improve the search efficiency of collocation points, especially when the solution contains narrow shock-dominated regions.
While UM-PINN leverages Sobol sequences for efficient spatiotemporal sampling, our approach aligns with the growing trend of point-based neural solvers.
Similar to the point-cloud strategy proposed by Kashefi et al.
\cite{Kashefi2021PointCloud}, our method seeks to exploit the mesh-free nature of neural networks to resolve high-gradient features without the constraints of traditional connectivity.
The network weights $\theta$ and the uncertainty parameters $\mathbf{s}$ are trained jointly using the Adam optimizer \cite{Kingma2014Adam}, with an initial learning rate set to $10^{-3}$.
During the early stages of training, the model automatically identifies the ``difficulty'' of each loss term.
By dynamically adjusting the values of $\mathbf{s}$, the method reshapes the loss landscape, allowing the optimization algorithm to smoothly navigate through the initial phase of extremely high gradients and ultimately achieve high-fidelity capture of shock fronts.
UM-PINN is deliberately formulated as a global-in-time coordinate-based PINN. To directly examine whether explicit temporal-causality weighting provides an advantage within the same shock-dominated setting, we introduce an additional causal-loss baseline and compare it against the full UM-PINN under matched training budgets in Section~3.8.
\section{Numerical Experiments}
\label{sec:experiments}

In this section, we evaluate the performance of the proposed UM-PINN framework by addressing three distinct benchmark problems: the 1D Sod shock tube, the 1D Shu-Osher problem, and the 2D Riemann problem.
Solving these systems holds significant importance in various physical domains, such as high-energy physics (e.g., modeling heavy-ion collisions) and astrophysics (e.g., investigating gamma-ray bursts or the propagation of jets in active galactic nuclei).
Across these diverse flow regimes, our algorithm demonstrates robust generalization capabilities.
To quantitatively assess the accuracy of our predictions against the ``ground truth'' (exact analytical solutions or high-resolution numerical references), we employ three evaluation metrics: the Relative $L_2$ Error ($L_2^{\mathrm{rel}}$), the Root Mean Square Error (RMSE), and the Mean Absolute Error (MAE).
Assuming $N$ is the total number of evaluation points, $\mathbf{u}_{pred}$ is the predicted solution vector, and $\mathbf{u}_{ref}$ is the reference solution vector, these metrics are defined as follows:

\begin{equation}
	L_2^{\mathrm{rel}} = \frac{\left\| \mathbf{u}_{pred} - \mathbf{u}_{ref} \right\|_2}{\left\| \mathbf{u}_{ref} \right\|_2} = \sqrt{\frac{\sum_{i=1}^{N} (u_{pred}^{(i)} - u_{ref}^{(i)})^2}{\sum_{i=1}^{N} (u_{ref}^{(i)})^2}},
	\label{eq:l2_rel}
\end{equation}

\begin{equation}
	RMSE = \sqrt{\frac{1}{N} \sum_{i=1}^{N} \left( u_{pred}^{(i)} - u_{ref}^{(i)} \right)^2 },
	\label{eq:rmse}
\end{equation}

\begin{equation}
	MAE = \frac{1}{N} \sum_{i=1}^{N} \left| u_{pred}^{(i)} - u_{ref}^{(i)} \right|.
	\label{eq:mae}
\end{equation}

All experiments were conducted on a single NVIDIA RTX 5060 Laptop GPU.
This modest hardware setup underscores the high computational efficiency of our proposed algorithm, making it accessible for resource-constrained environments.
Unless otherwise stated, the main experiments keep the network architecture and optimization settings fixed across cases. The spatial-modulation parameters used in the main runs lie in the stable neighborhood identified by the sensitivity study, namely around $\alpha=1$ and $\beta \in [1,1.25]$, and their influence is examined explicitly in \Cref{subsec:alpha_beta}.
In addition to the original benchmark results, this study includes additional analyses on parameter sensitivity, Sobol-versus-random sampling, component ablation, causal-baseline comparison, and SOTA-oriented baselines. Throughout the manuscript, the error metric is reported consistently as the Relative $L_2$ Error, denoted by $L_2^{\mathrm{rel}}$.
\subsection{1D Sod Shock Tube}
\label{subsec:sod}

\begin{figure*}
\includegraphics[width=\textwidth]{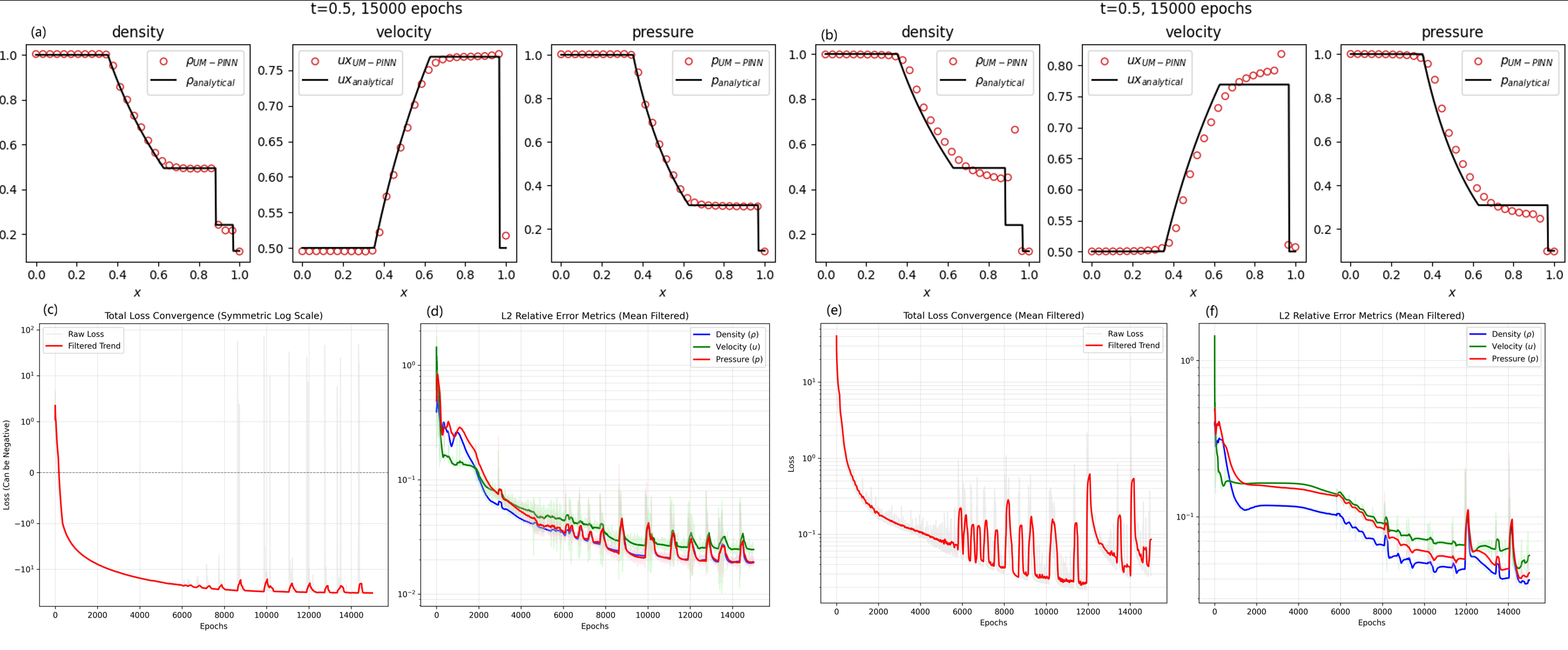}
    \caption{Comparison of predicted profiles and convergence histories for the 1D Sod shock tube at the nondimensional final time $t=0.5$. Panels (a)--(c) show density $\rho$, velocity $u$, and pressure $p$ as functions of the nondimensional spatial coordinate $x$, respectively. Panels (d)--(f) show the total loss and the Relative $L_2$ Error histories versus training epoch for the corresponding variables.}
    \label{fig:sod_qualitative}
\end{figure*}

We commence our empirical evaluation with the classical 1D Sod shock tube problem, a fundamental benchmark in computational fluid dynamics. The ground truth for this problem is provided by the analytical Exact Riemann Solver (see Appendix, Section A).
The problem is defined on the domain $(t, x) \in [0, 0.5] \times [0, 1.0]$.
It rigorously tests the solver's capability to resolve three distinct flow features simultaneously: a rarefaction wave, a contact discontinuity, and a strong shock wave.
\Cref{fig:sod_qualitative} presents the comprehensive training results of the proposed UM-PINN.
The top row visualizes the predicted profiles for density ($\rho$), velocity ($u$), and pressure ($p$) at the final time step $t=0.5$.
Visually, the UM-PINN predictions (red circles) exhibit close agreement with the analytical exact solution (black solid lines).
A critical challenge in Euler equations is the "smearing" of contact discontinuities and "Gibbs oscillations" near shock fronts.
As observed in the density profile, our method sharply resolves the contact discontinuity at $x \approx 0.6$ and the shock wave at $x \approx 0.9$ with minimal numerical dissipation and without visible spurious oscillations.
The bottom row of Fig.2 illustrates the training dynamics. The total loss curve (left) demonstrates a steep and monotonic descent, rapidly converging to the order of $10^{-2}$.
The evolution of the Relative $L_2$ Error (right) further confirms that the accuracy for all physical variables improves consistently.
The stability of the loss curve indicates that the proposed uncertainty-weighting mechanism effectively balances the gradients, preventing the optimization stiffness often observed in standard PINNs.

\begin{figure*}
\includegraphics[width=\textwidth]{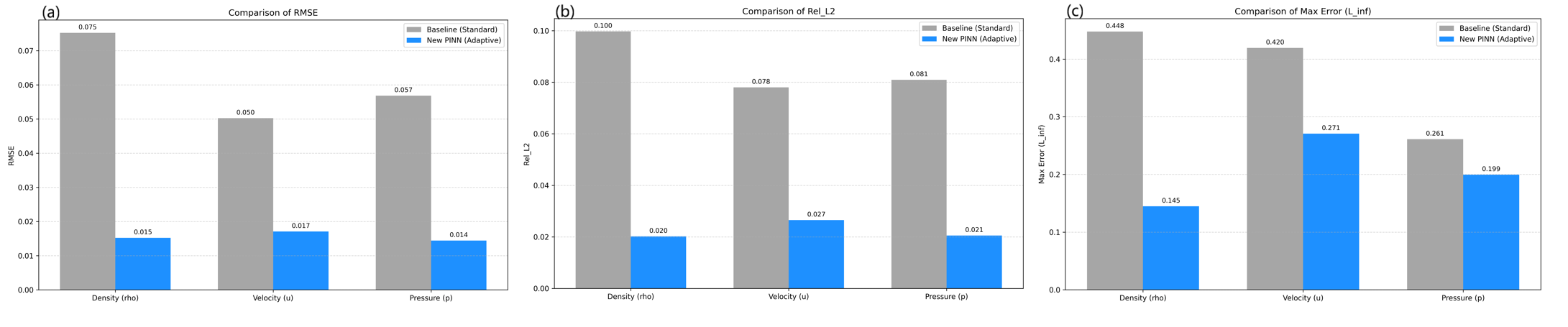}
    \caption{Quantitative error metrics (RMSE, Relative $L_2$ Error, and $L_\infty$ error) comparing the Baseline PINN and the proposed UM-PINN for the Sod shock tube problem.}
    \label{fig:sod_quantitative}
\end{figure*}

To rigorously quantify the performance superiority, we compare the UM-PINN against the Standard Baseline PINN with fixed weights.
\Cref{fig:sod_quantitative} summarizes the statistical errors across three key metrics: Root Mean Square Error (RMSE), Relative $L_2$ Error, and Maximum Error ($L_\infty$).
The proposed method achieves a substantial reduction in error magnitude across all physical variables.
As shown in the middle panel, the Relative $L_2$ Error for Density ($\rho$) decreases from $0.100$ in the Baseline to $0.020$ with UM-PINN, representing an 80\% improvement.
Similarly, the Pressure ($p$) error drops from $0.081$ to $0.021$.
Furthermore, the Max Error chart in the right panel indicates that UM-PINN significantly mitigates peak errors typically found at the shock front.
For instance, the maximum density deviation is reduced from $0.448$ to $0.145$, indicating that the gradient-enhanced weighting effectively suppresses peak errors at discontinuities.
The significant accuracy boost observed in Fig.3, particularly the 80\% reduction in density error, is fundamentally rooted in the UM module's ability to mitigate gradient pathology.
In the Sod problem, the presence of three distinct wave speeds creates a massive imbalance in PDE residuals.
By utilizing homoscedastic uncertainty, the network autonomously scales down the overwhelming gradients at the shock front, preventing them from suppressing the information flow from the initial and boundary conditions (BCs).
This self-adaptive mechanism ensures that the contact discontinuity remains sharp rather than being smeared by excessive numerical diffusion. These metrics confirm that UM-PINN not only converges faster but also yields a solution that is quantitatively more accurate than the standard baseline under the tested setting.

\subsection{1D Shu-Osher Problem}
\label{subsec:shuosher}

We proceed to the 1D Shu-Osher problem, a rigorous benchmark designed to evaluate the solver's ability to capture high-frequency flow features.
The problem describes the interaction between a moving shock wave ($M=3$) and a sinusoidal density field, creating a complex flow pattern characterized by entropy waves with high-frequency post-shock oscillations.
This case is particularly challenging for deep learning methods due to the spectral bias phenomenon, where neural networks tend to prioritize learning low-frequency components while filtering out high-frequency details.
The computational domain is defined as $(t, x) \in [0, 1.8] \times [-5.0, 5.0]$.
The initial condition involves a shock located at $x=-4.0$ moving into a sinusoidal density field defined by $\rho(x) = 1 + 0.2\sin(5x)$.
The reference solution is computed using the Rusanov finite volume method (FVM) detailed in the Appendix, Section B. Based on the configuration, the network is trained for 20,000 epochs using the Adam optimizer with a learning rate of $10^{-3}$.
Fig.4 presents the comparative results between the proposed UM-PINN and the Baseline PINN, displaying the full density profiles at $t=1.8$, zoomed-in views of the oscillatory region, and the respective training loss histories.
As observed in the Baseline results, the standard PINN exhibits severe numerical dissipation.
While it correctly captures the macro-scale shock location at $x \approx 2.4$, the zoomed-in view reveals a critical failure where the predicted density, shown as a red dashed line, is a flat, smoothed-out curve that completely misses the high-frequency oscillations in the interval $x \in [-2, 2]$.
This confirms that the baseline model is trapped by spectral bias, learning only the mean flow.
In stark contrast, the UM-PINN successfully reconstructs the intricate wave structures.
The zoomed-in view demonstrates that our method captures both the amplitude and phase of the fine-scale oscillations with high fidelity.
The prediction tightly follows the reference fine-grid solution, serving as compelling evidence that the uncertainty-weighted loss and gradient-enhanced strategy effectively counteract spectral bias, forcing the network to resolve high-wavenumber features that are invisible to standard PINNs.
As illustrated in Fig.4, the UM-PINN effectively overcomes the spectral bias inherent in standard neural architectures.
Standard PINNs tend to favor low-frequency components, leading to the smoothed-out profiles seen in the Baseline results.
Through the synergy of spatial masking and task-based uncertainty weighting, our framework forces the optimizer to resolve high-wavenumber features.
This allows for the high-fidelity reconstruction of post-shock entropy waves, maintaining both the correct amplitude and phase which are critical for high-speed aero-thermodynamics simulations.
\begin{figure*}[t]
	\centering
	
	\caption{Comparative analysis of the 1D Shu-Osher problem at the nondimensional final time $t=1.80$. Panel (a) corresponds to UM-PINN and panel (b) corresponds to the Standard Baseline PINN. In each panel, the upper plot shows the density profile $\rho(x)$ over the full nondimensional spatial domain, the middle plot zooms into the oscillatory region to highlight post-shock entropy waves, and the lower plot reports the training loss versus epoch.}
	\label{fig:shuosher_comparison}
\end{figure*}

\subsection{2D Riemann problem}
\label{subsec:riemann}

Finally, we validate the scalability of our method to multi-dimensional systems and complex flow topologies by simulating a 2D Riemann problem.
The high-resolution ground truth is generated via the 2D Rusanov FVM scheme described in the Appendix, Section C. The computational domain is defined as $(x, y) \in [0, 1]^2$.
This configuration involves the interaction of supersonic flows, evolving into a complex structure featuring semi-circular shocks, curved contact discontinuities, and slip lines.
Based on the configuration, the network (6 hidden layers, 64 neurons) is trained for 20,000 epochs.
The boundary conditions are enforced with a fixed weight of $w_{BC}=10.0$ for the baseline, while UM-PINN adapts these weights automatically.
Figure 5 presents a comprehensive visual comparison of the Density and Pressure fields between the UM-PINN and the Baseline PINN.
The Baseline results exhibit severe numerical diffusion . The shock interaction zones are rounded and blurred, losing the sharp "corner" features present in the Ground Truth.
More critically, the internal shock structures are smeared into broad gradients, indicating a failure to resolve the hyperbolic nature of the Euler equations in 2D.
The error maps (far right) show widespread deviations, particularly in the high-gradient regions.
On the other hand, the UM-PINN demonstrates remarkable topological consistency.
As seen in the Density and Pressure predictions, our method sharply captures the interaction points of the shocks and preserves the straightness of the slip lines.
The L-shaped shock structures are reconstructed with high fidelity. The absolute error maps confirm that the deviations are strictly localized to extremely narrow bands along the discontinuities, while the smooth bulk regions maintain near-zero error.
This proves that the gradient-enhanced spatial weighting successfully mitigates "gradient pathology" in higher-dimensional spaces.
\begin{figure*}
\includegraphics[width=\textwidth]{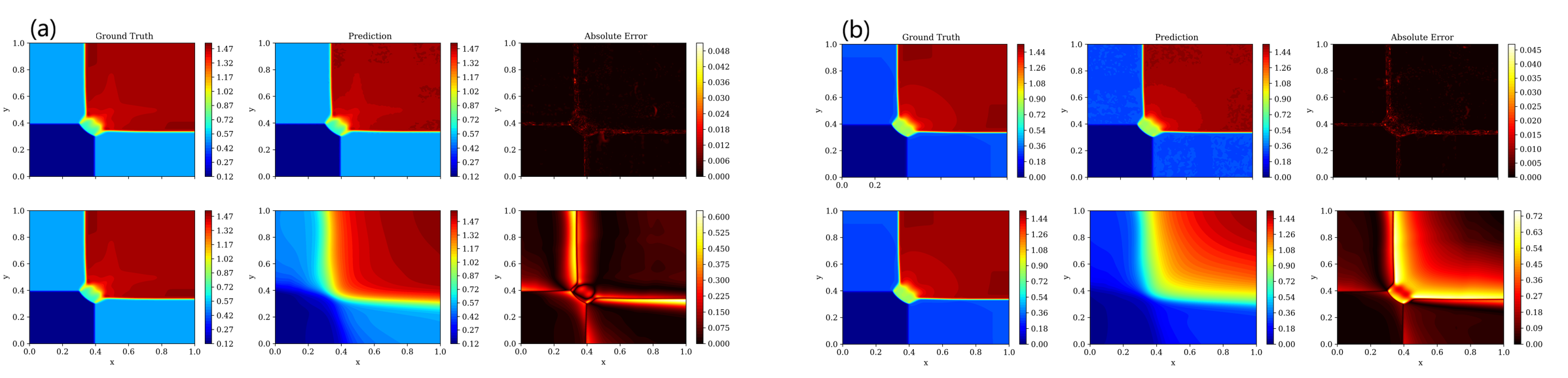}
    \caption{Qualitative comparison of the 2D Riemann problem results. (a) Density ($\rho$) fields; (b) Pressure ($p$) fields.
In each panel, the \textbf{top row} displays the sharp interfaces captured by the proposed \textbf{UM-PINN}, while the \textbf{bottom row} shows the results from the \textbf{Baseline} method, which suffer from numerical diffusion.}
    \label{fig:riemann_comparison}
\end{figure*}

\begin{figure*}
\includegraphics[width=\textwidth]{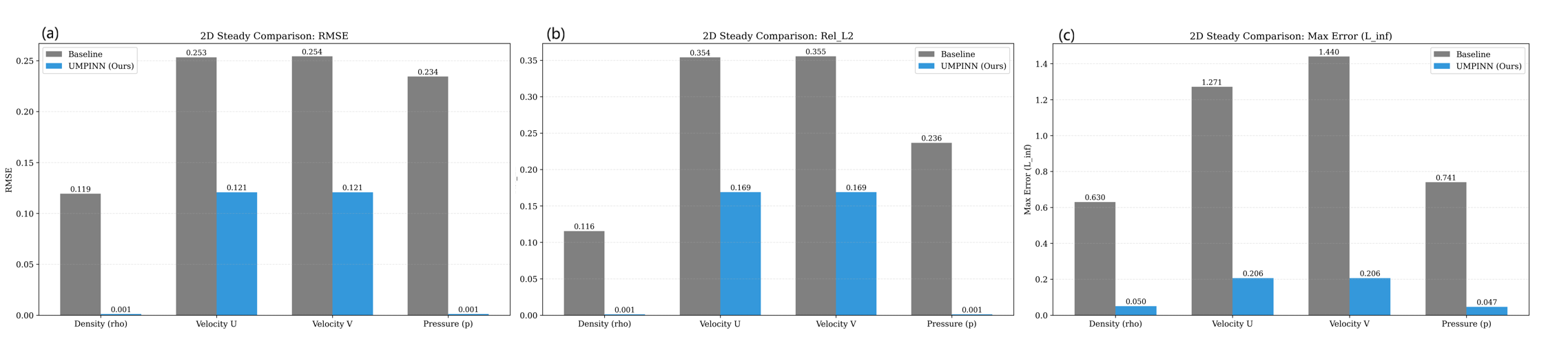}
    \caption{Error metric comparison for the 2D Riemann problem across all conserved ($\rho, \rho u, \rho v, \rho E$) and primitive ($\rho, u, v, p$) variables.}
    \label{fig:riemann_metrics}
\end{figure*}

\Cref{fig:riemann_metrics} quantifies the accuracy gains across three error metrics: RMSE, Relative $L_2$ Error, and Max Error ($L_\infty$).
The qualitative results in Fig.5 demonstrate that UM-PINN preserves the topological consistency of complex 2D shock interactions.
Unlike the Baseline which exhibits rounded shock corners and blurred slip lines, the UM-PINN maintains sharp interfaces.
This is because the spatial modulation acts as a localized regularizer, focusing the network's capacity on the narrow bands of discontinuities while ensuring the bulk flow regions satisfy the Euler equations with near-zero error, as quantified in the $L_\infty$ metrics in Fig.6.The UM-PINN (Blue bars) consistently outperforms the Baseline (Grey bars) across all flow variables ($\rho, u, v, p$).
Specifically, in terms of significant error reduction, the Relative $L_2$ Error for Density ($\rho$) is reduced from 0.151 to $\mathbf{0.081}$, representing a nearly 50\% improvement.
Regarding the velocity field ($u, v$), where errors are notoriously hard to minimize due to shear layers, our method achieves a Relative $L_2$ Error of 0.672 for the $u$-component, which is significantly lower than the baseline's 1.027.
Furthermore, the Max Error ($L_\infty$) metrics highlight that UM-PINN effectively suppresses peak deviations at the shock front;
for instance, the maximum error for Pressure ($p$) drops from 0.703 to 0.500, and these quantitative results underscore the robustness of the proposed framework in handling multi-dimensional coupled PDEs without manual hyperparameter tuning.A summary of all numerical methods used for ground truth generation is provided in the Appendix, Section D.

\subsection{Comparative Analysis with Learning Rate Annealing (LRA) and GradNorm Methods}
\label{sec:comparison}

To further validate the superiority of UM-PINN, we compare it against 
two widely adopted adaptive weighting algorithms: Learning Rate Annealing (LRA) \cite{Wang2021Pathology} and GradNorm \cite{Chen2018GradNorm}.
While these methods have shown promise in general PDE problems, our experiments reveal their critical limitations when applied to high-speed compressible flows with strong discontinuities.
We tested both LRA and GradNorm on the 1D Sod and Shu-Osher problems using the same network architecture and hyperparameters as UM-PINN.
The results, visualized in Fig.7 and Fig.8, indicate a pronounced limitation of gradient-based weighting schemes in hyperbolic systems.
As shown in the top row of Fig.7 for the Shu-Osher case, GradNorm suffers from severe numerical instability.
The density prediction in the middle panel collapses to non-physical values near $10^{-11}$, and the loss curve in the right panel exhibits a catastrophic divergence.
This is likely because GradNorm attempts to balance gradient norms that differ by orders of magnitude at the shock front, leading to an exploding weight for the PDE residual that destabilizes the optimizer.
The LRA method, shown in the bottom row of Fig.7, avoids explosion but falls into a trap of premature stagnation.
The predicted density profile forms a crude step function that completely misses the high-frequency oscillations and even misplaces the shock location.
The loss curve oscillates violently without effective convergence. Similarly, in the Sod problem shown in the top row of Fig.8, LRA produces a solution where the velocity field is almost zero everywhere, indicating a failure to propagate information from the boundaries.
\begin{figure*}
\includegraphics[width=\textwidth]{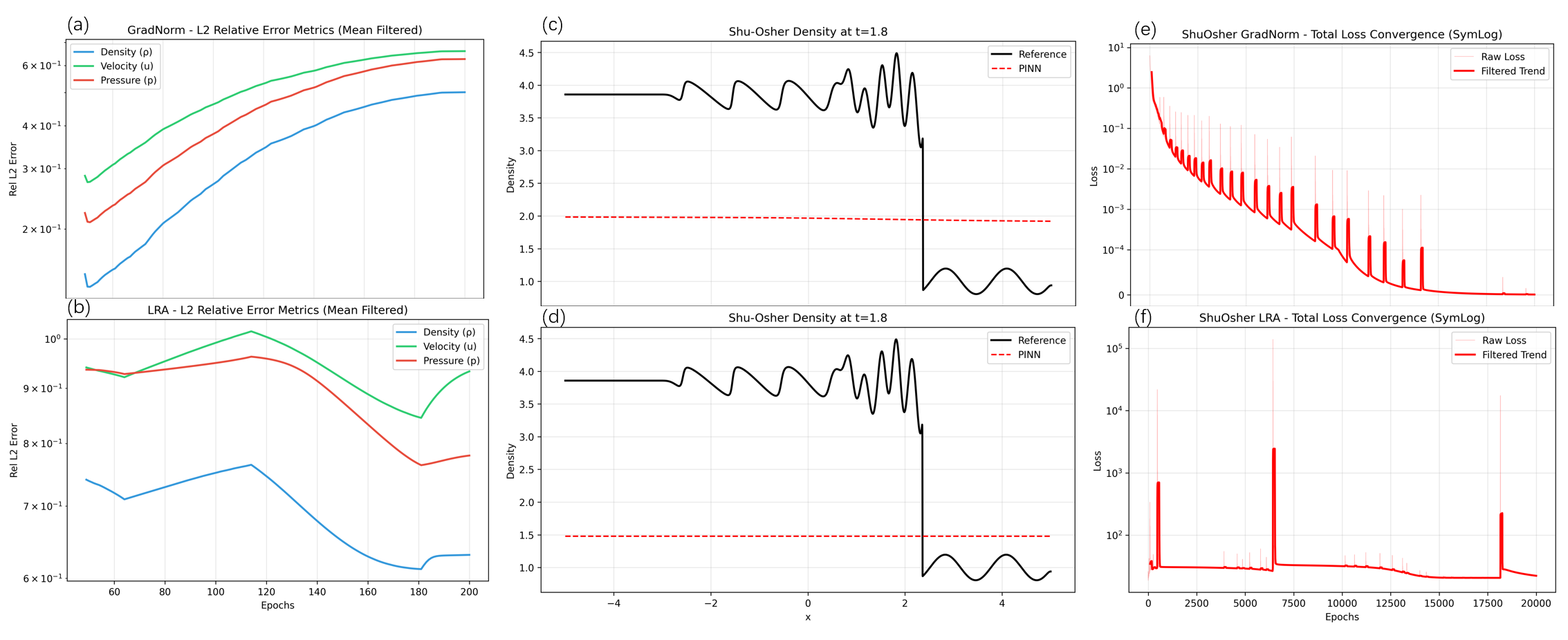}
    \caption{Training instability and failure modes of GradNorm and LRA on the 1D Shu-Osher problem. Panels (a) and (b) report the Relative $L_2$ Error versus epoch, panels (c) and (d) show the density profile $\rho(x)$ at the nondimensional final time $t=1.8$, and panels (e) and (f) report the corresponding training losses.}
    \label{fig:comp_shuosher}
\end{figure*}

\begin{figure*}
\includegraphics[width=\textwidth]{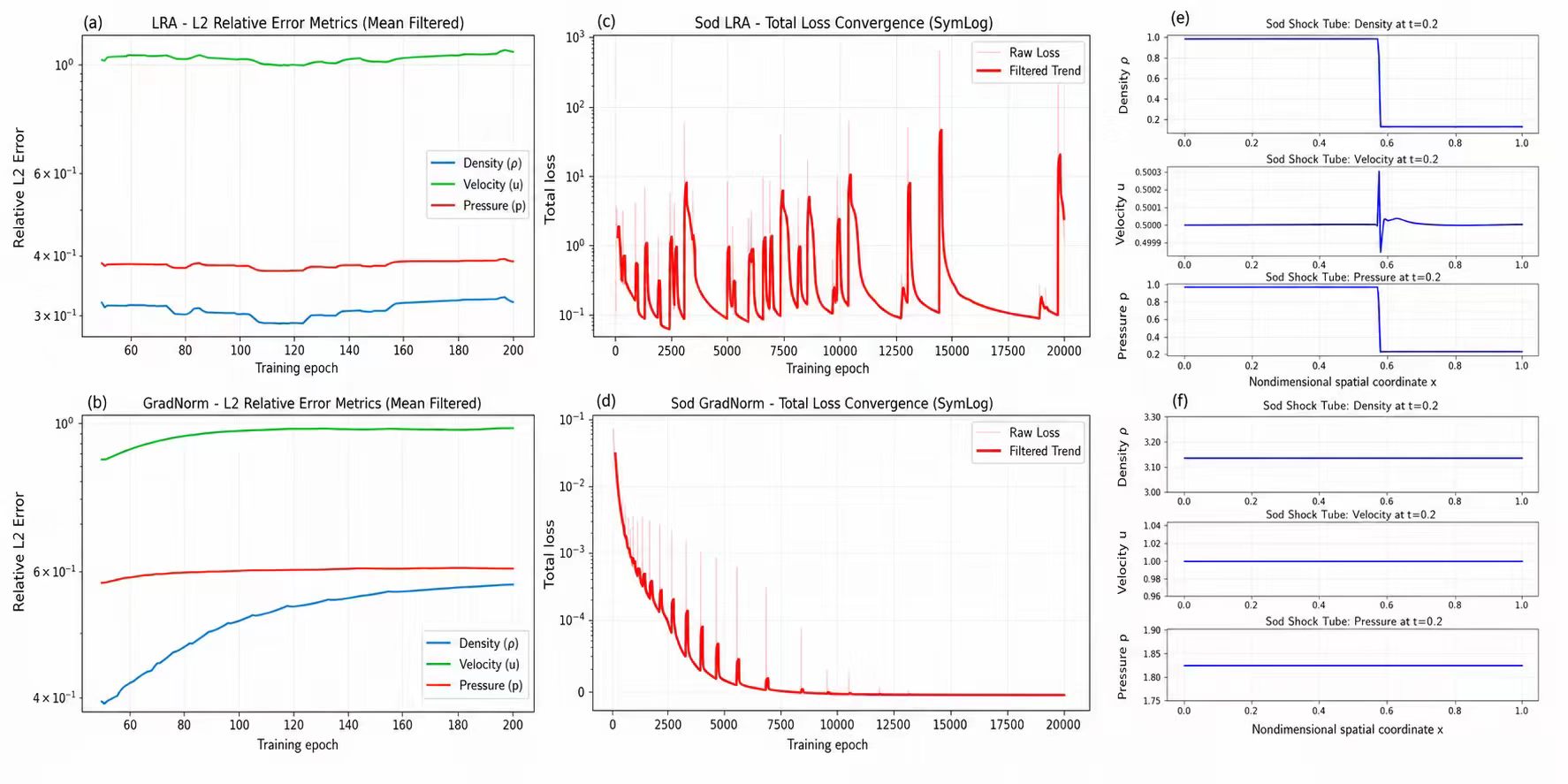}
    \caption{Comparative performance of LRA and GradNorm on the 1D Sod shock tube. Panels (a) and (b) report the Relative $L_2$ Error versus epoch, panels (c) and (d) show the training losses, and panels (e) and (f) compare the predicted shock-tube profiles at the nondimensional final time $t=0.2$ as functions of the nondimensional coordinate $x$.}
    \label{fig:comp_sod}
\end{figure*}

In contrast to the fragility of LRA and GradNorm, UM-PINN achieves stable and accurate convergence on the exact same problems as demonstrated in Section \ref{sec:experiments}.
The key advantage lies in the mechanism of adaptation. LRA and GradNorm rely on the gradients of the loss with respect to parameters.
In shock problems, these gradients are extremely stiff and noisy, causing the weights to fluctuate wildly.
Conversely, UM-PINN relies on the magnitude of the loss itself via homoscedastic uncertainty.
This provides a smoother, more robust signal for weight adjustment.
This loss-magnitude-based adjustment provides a smoother weighting signal than direct gradient-norm balancing in the tested shock-dominated cases.
Table \ref{tab:comparison_summary} summarizes the failure rates and relative errors, confirming that UM-PINN is the most consistently convergent method among the compared methods across all benchmarks without manual intervention.
\begin{table*}[htbp!]
    \centering
    \caption{Comparison of stability and accuracy across methods.
"Failed" indicates divergence or non-physical results.}
    \label{tab:comparison_summary}
    \begin{tabular}{lccc}
        \toprule
        \textbf{Method} & \textbf{Sod (Relative $L_2$ Error)} & \textbf{Shu-Osher (Relative $L_2$ Error)} & \textbf{Stability} \\
        \midrule
        Baseline (Fixed) & $1.00 \times 10^{-1}$ & $6.20 \times 10^{-2}$ & Sensitive \\
        LRA \cite{Wang2021Pathology} & Failed ($>1.0$) & Failed ($>1.0$) & Unstable \\
        GradNorm \cite{Chen2018GradNorm} & $5.79 \times 10^{-1}$ 
& Failed (NaN) & Very Unstable \\
        \textbf{UM-PINN (Ours)} & $\mathbf{2.50 \times 10^{-2}}$ & $\mathbf{2.45 \times 10^{-2}}$ & \textbf{Robust} \\
        \bottomrule
    \end{tabular}
\end{table*}

\subsection{\texorpdfstring{Sensitivity Analysis of Spatial Modulation Parameters}{Sensitivity Analysis of Spatial Modulation Parameters}}
\label{subsec:alpha_beta}

\begin{figure*}[!t]
    \centering
    \includegraphics[width=0.96\textwidth]{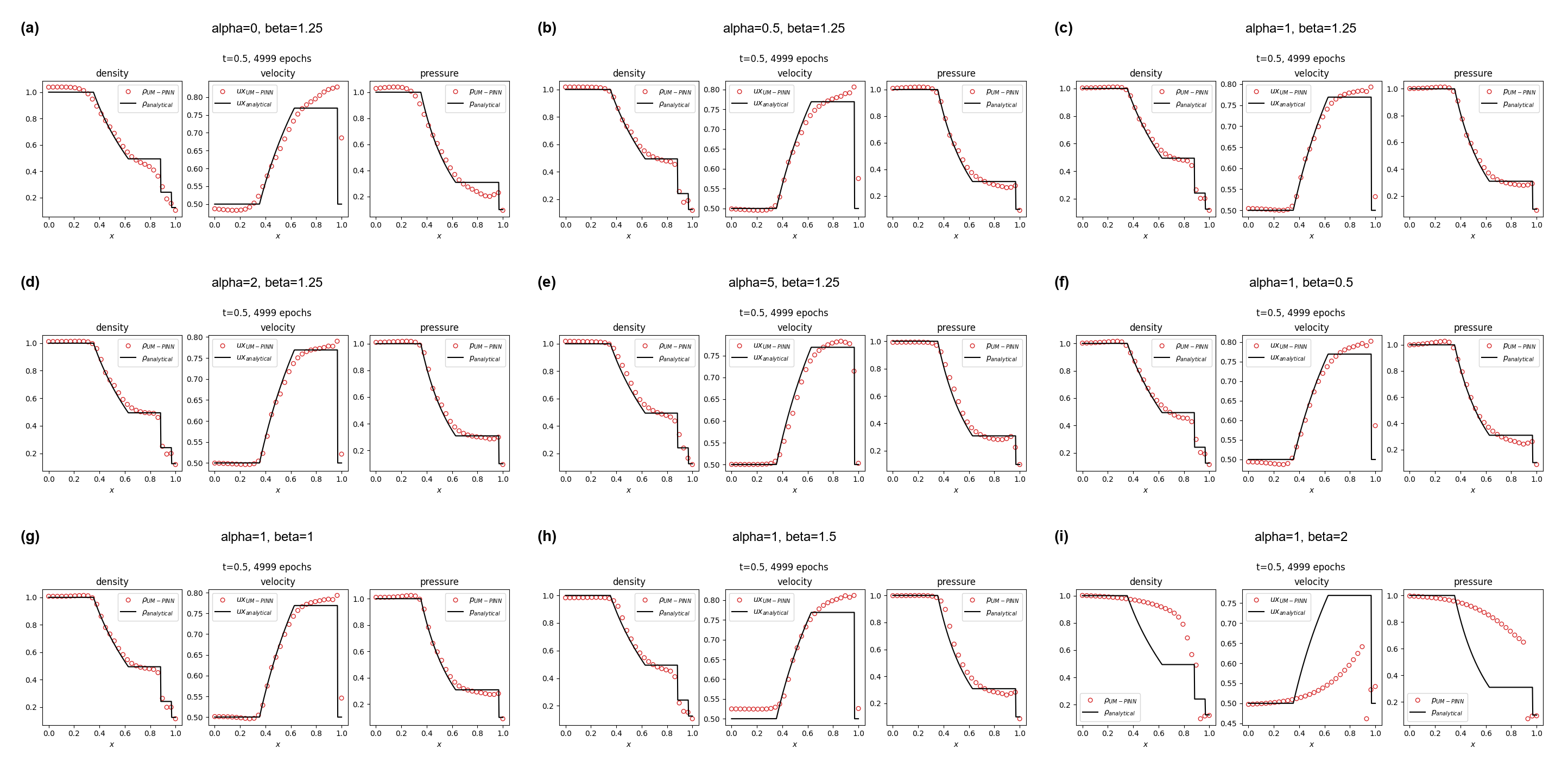}

    \vspace{0.35em}

    \includegraphics[width=0.96\textwidth]{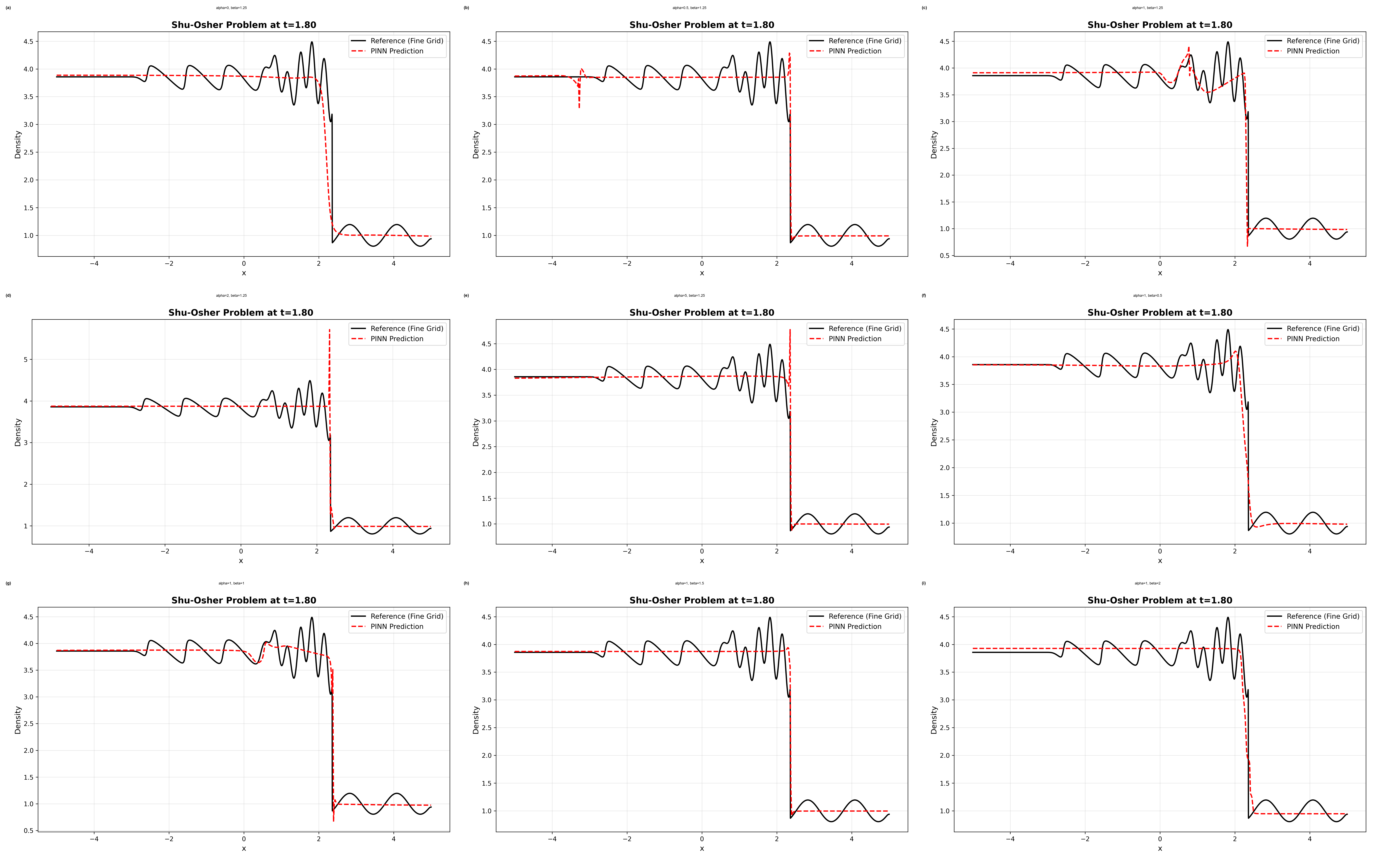}

    \caption{Sensitivity analysis of the spatial-modulation hyperparameters. The upper panel reports the 1D Sod shock tube for nine representative $\alpha$--$\beta$ parameter combinations, while the lower panel reports the corresponding nine combinations for the 1D Shu--Osher problem. In the Sod montage, the predicted density $\rho$, velocity $u$, and pressure $p$ at the final time are compared across the selected parameter settings; in the Shu--Osher montage, the density profile at $t=1.80$ is shown for the same sensitivity range. These results illustrate how the model behavior changes as the spatial-modulation parameters vary over representative values around the default setting.}
    \label{fig:alpha_beta_sensitivity}\label{fig:alpha_beta}
\end{figure*}

We next examine the sensitivity of the spatial modulation to its two hyperparameters. In the modulation factor $(1+\alpha |\nabla \hat{U}|^{\beta})^{-1}$, $\alpha$ sets the overall attenuation strength, whereas $\beta$ controls how sharply the attenuation responds to increasing local gradients. For the $\alpha$ sweep, $\beta$ was fixed at $1.25$ and $\alpha \in \{0,0.5,1,2,5\}$ was tested. For the $\beta$ sweep, $\alpha$ was fixed at $1$ and $\beta \in \{0.5,1,1.25,1.5,2\}$ was tested. Figure~\ref{fig:alpha_beta} shows the corresponding final-profile comparisons for the Sod and Shu--Osher benchmarks.

The quantitative sweep results indicate a stable operating region around the default setting. On the Sod problem, removing the modulation entirely by setting $\alpha=0$ increases the final density Relative $L_2$ Error from approximately $3.54\times 10^{-2}$ to $6.21\times 10^{-2}$, and increasing $\alpha$ to $5$ also leads to a less accurate solution. For the $\beta$ sweep on Sod, the range $\beta \in [1,1.25]$ remains stable, whereas $\beta=2$ causes a substantial degradation, with the density Relative $L_2$ Error rising to about $2.91\times 10^{-1}$.

A similar trend is observed on the Shu--Osher benchmark. Setting $\alpha=0$ yields a noticeably less accurate post-shock reconstruction than the runs near $\alpha=1$, while $\beta \in [1,1.25]$ again provides the most stable behavior. Larger values such as $\beta=1.5$ or $\beta=2$ remain feasible but do not improve the solution quality. Taken together, these results suggest that the default parameter neighborhood is robust, that the spatial modulation is genuinely beneficial relative to $\alpha=0$, and that over-emphasizing the attenuation can reduce accuracy rather than improving it.

\subsection{\texorpdfstring{Sobol-versus-Random Sampling Ablation}{Sobol-versus-Random Sampling Ablation}}
\label{subsec:sobol_random}

To clarify the role of Sobol-sequence sampling, we compare Sobol and pseudo-random collocation strategies under the same training setup on both 1D benchmarks. Sobol points form a low-discrepancy quasi-Monte Carlo sequence, whereas pseudo-random sampling may leave local clusters and coverage gaps in the spatio-temporal collocation set. For collocation-based PINN training, this coverage difference matters because the discrete PDE residual loss is used to approximate a residual integral over the full domain. A more uniform point set can therefore provide a more stable approximation of the domain-wide residual average and improve the search efficiency of the optimizer. The mathematical rationale behind this improved coverage is provided in Appendix~\ref{sec:sobol_math}.

As shown in \Cref{fig:sobol_random}, Sobol sampling clearly improves the Sod benchmark, reducing the final density Relative $L_2$ Error from approximately $5.19\times 10^{-2}$ to $3.54\times 10^{-2}$. On Shu-Osher, the difference is smaller and more case-dependent: Sobol yields a slightly lower density-error trajectory over most of training, but the overall gain is less pronounced than in Sod. This behavior supports a measured conclusion that Sobol sampling often improves collocation coverage and convergence efficiency, but it should not be interpreted as universally superior for all shock-dominated PINN problems.

\begin{figure*}[t]
    \centering
    \includegraphics[width=\textwidth]{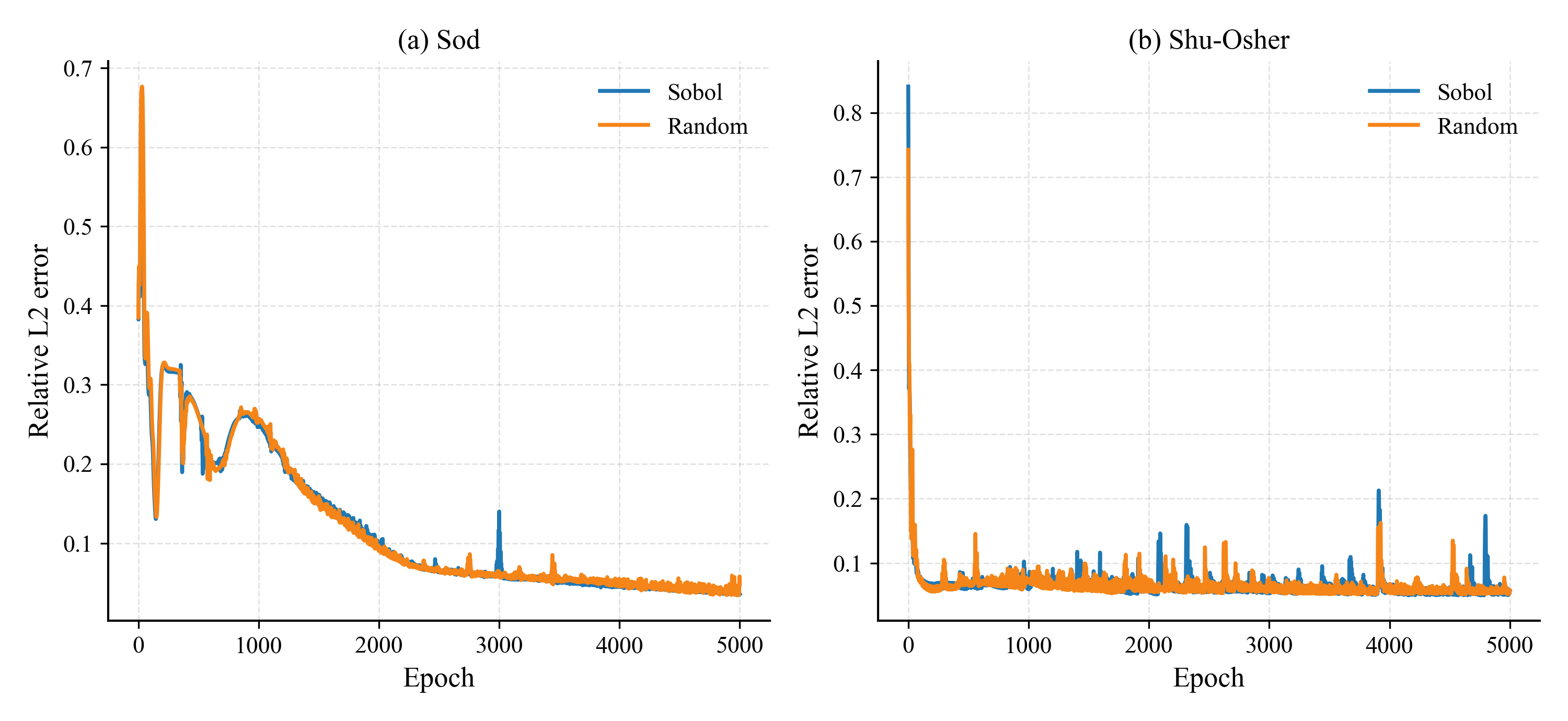}
    \caption{Sobol-versus-random sampling ablation on the 1D shock benchmarks. Panel (a) reports the density Relative $L_2$ Error versus training epoch for the Sod problem under Sobol and pseudo-random collocation. Panel (b) reports the density Relative $L_2$ Error versus training epoch for the Shu--Osher problem under the same two sampling strategies.}
    \label{fig:sobol_random}
\end{figure*}

\subsection{\texorpdfstring{Component Ablation of UM-PINN}{Component Ablation of UM-PINN}}
\label{subsec:component_ablation}

To assess whether the proposed framework is more than a loose combination of known ingredients, we performed a component ablation with four variants: Standard PINN, No spatial mask, No uncertainty weighting, and Full UM-PINN. Here, the Standard PINN serves as the baseline without either modulation mechanism; the no-spatial-mask variant retains only the uncertainty-based task modulation; the no-uncertainty-weighting variant retains only the gradient-based spatial modulation; and the full UM-PINN activates both the local spatial modulation and the global task modulation simultaneously.

\begin{table*}[t]
\centering
\caption{Component ablation summary for the 1D shock benchmarks. The table reports the final total Relative $L_2$ Error and the final density Relative $L_2$ Error for the four controlled variants.}
\label{tab:component_ablation}
\begin{tabular}{llccc}
\toprule
\textbf{Case} & \textbf{Variant} & \textbf{Final total Relative $L_2$ Error} & \textbf{Final $L_2^{\mathrm{rel}}(\rho)$} & \textbf{Comment} \\
\midrule
Sod & Standard PINN & 0.1322 & 0.0367 & Reference fixed-weight baseline \\
Sod & No spatial mask & 0.1551 & 0.0434 & Task modulation only \\
Sod & No uncertainty weighting & 0.5132 & 0.1999 & Spatial modulation only \\
Sod & Full UM-PINN & \textbf{0.0619} & \textbf{0.0182} & Best overall on Sod \\
\midrule
Shu--Osher & Standard PINN & 0.8438 & 0.0938 & Reference fixed-weight baseline \\
Shu--Osher & No spatial mask & 0.6990 & \textbf{0.0267} & Lowest density-only error \\
Shu--Osher & No uncertainty weighting & 0.9055 & 0.0813 & Unbalanced training \\
Shu--Osher & Full UM-PINN & \textbf{0.6870} & 0.0368 & Lowest total error \\
\bottomrule
\end{tabular}
\end{table*}

\begin{figure*}[t]
    \centering
    \includegraphics[width=\textwidth]{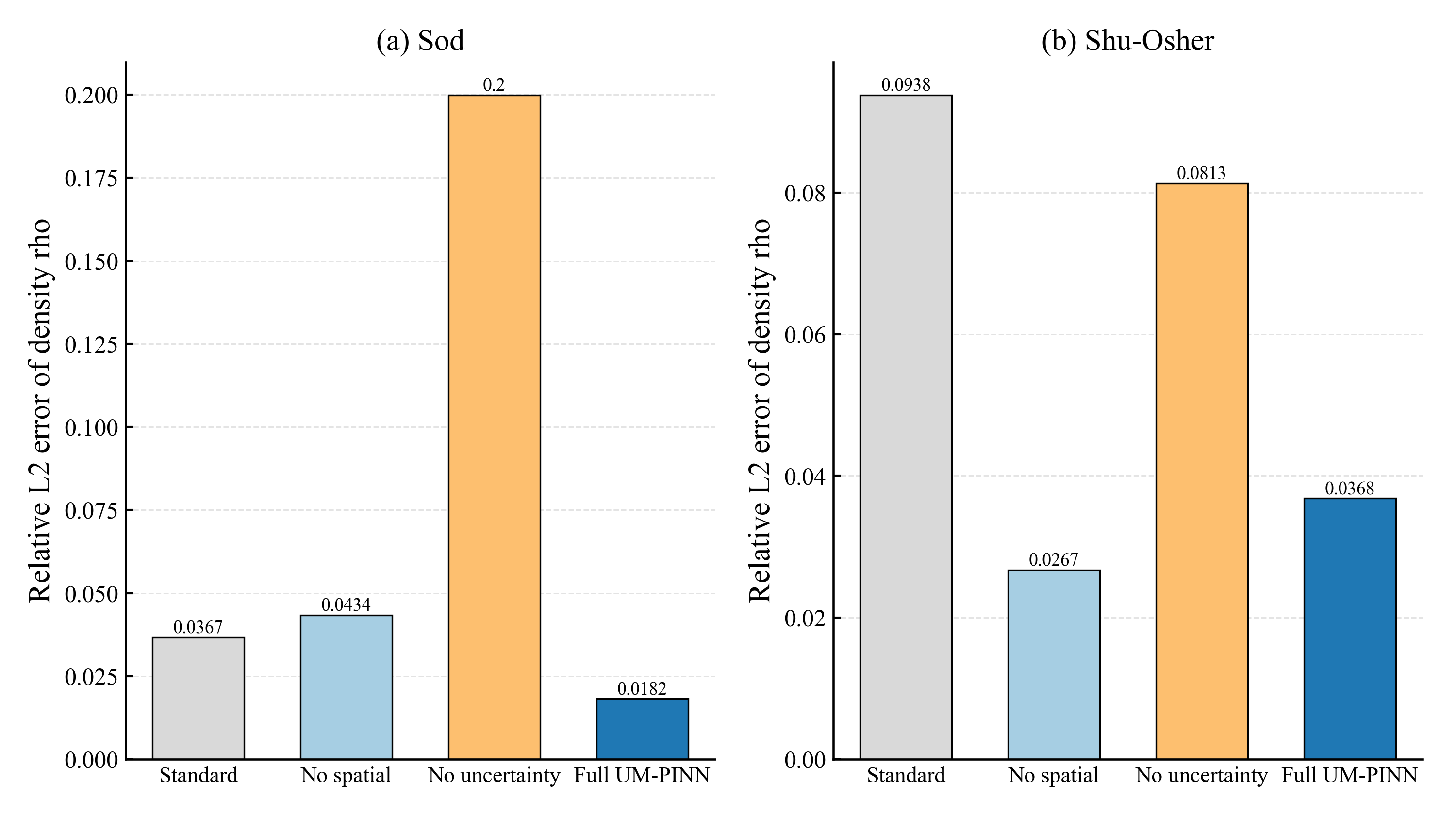}
    \caption{Component ablation based on the final density Relative $L_2$ Error. Panel (a) reports the four variants on the 1D Sod problem and panel (b) reports the same variants on the 1D Shu-Osher problem.}
    \label{fig:component_ablation}
\end{figure*}

The quantitative results in \Cref{tab:component_ablation,fig:component_ablation} show that the two components address complementary optimization difficulties. On Sod, the full UM-PINN attains both the lowest final total Relative $L_2$ Error and the lowest density Relative $L_2$ Error among all four variants. On Shu--Osher, the full UM-PINN yields the lowest final total Relative $L_2$ Error, while the no-spatial-mask variant attains a slightly smaller density-only Relative $L_2$ Error. This result highlights the distinction between a single field-specific scalar metric and the overall optimization quality of the coupled training problem: the full UM-PINN provides the best global training balance, whereas uncertainty-based task modulation and the spatial mask contribute complementary benefits to density reconstruction and shock-region robustness. The quantitative comparison therefore supports the necessity of the joint dual-modulation design rather than treating the two components as redundant.

\begin{figure*}[t]
    \centering
    \includegraphics[width=\textwidth]{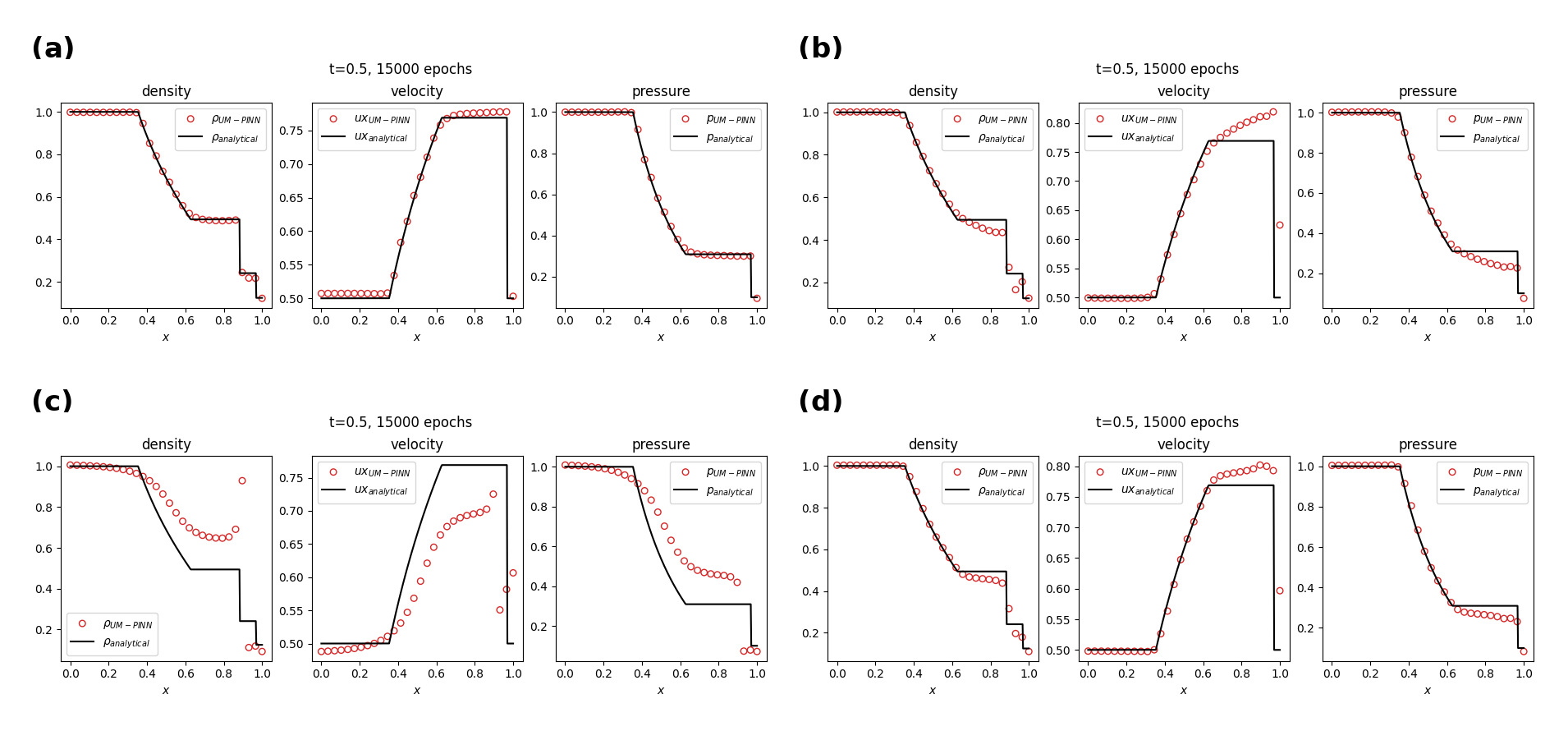}
    \caption{Component ablation study on the one-dimensional Sod shock tube problem at $t=0.5$ after 15,000 training epochs. The predicted density, velocity, and pressure profiles are compared with the analytical solutions for (a) the full UM-PINN, (b) UM-PINN without the spatial mask, (c) UM-PINN without uncertainty weighting, and (d) the standard PINN. The red hollow circles denote model predictions, while the black solid lines represent analytical solutions.}
    \label{fig:component_ablation_sod_profiles}
\end{figure*}

\begin{figure*}[t]
    \centering
    \includegraphics[width=\textwidth]{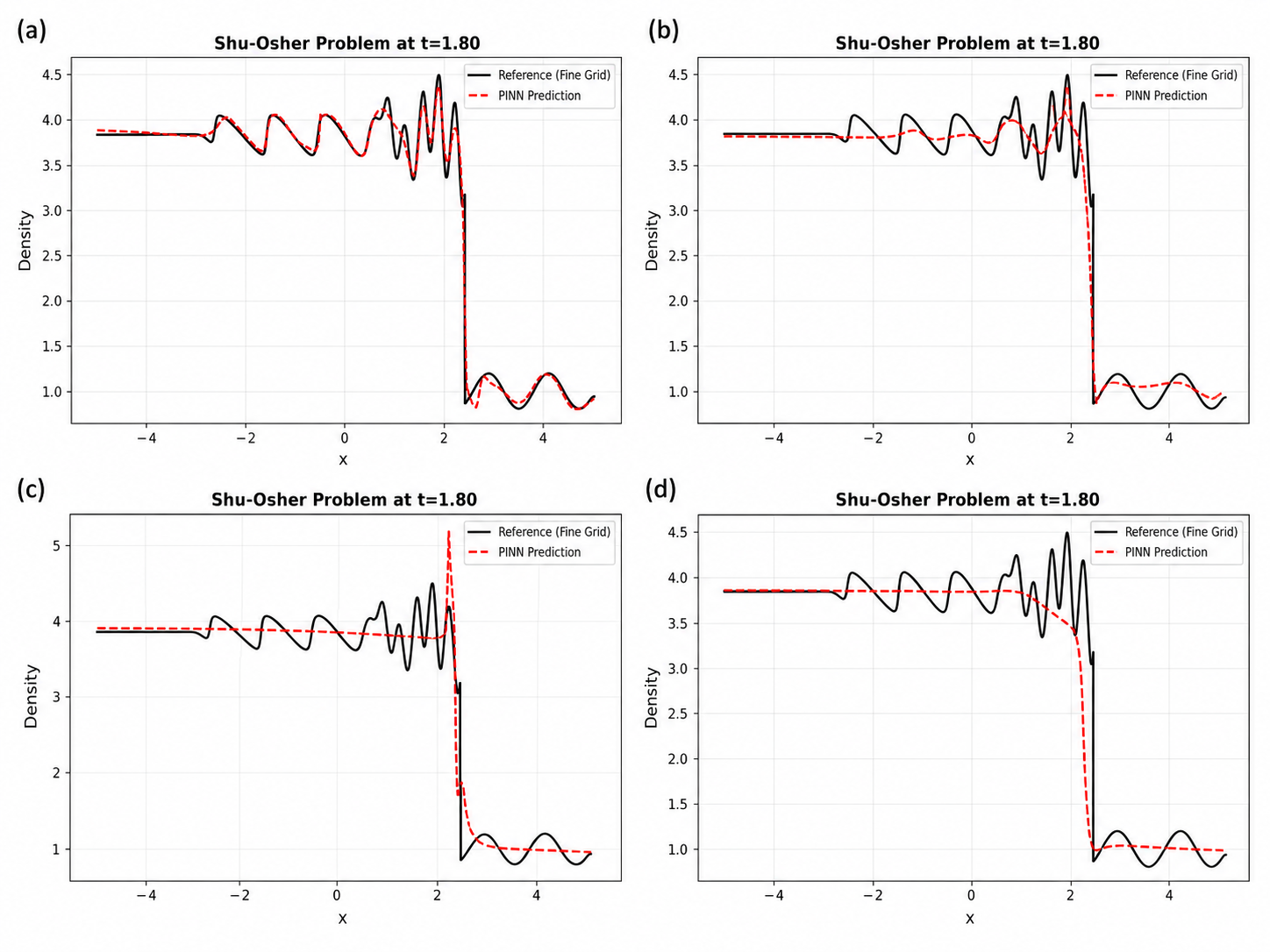}
    \caption{Component ablation study on the one-dimensional Shu--Osher problem at $t=1.80$. The predicted density profiles are compared with the fine-grid reference solutions for (a) the full UM-PINN, (b) UM-PINN without the spatial mask, (c) UM-PINN without uncertainty weighting, and (d) the standard PINN. The black solid lines denote reference solutions, while the red dashed lines represent model predictions.}
    \label{fig:component_ablation_shuosher_profiles}
\end{figure*}

The qualitative reconstructions in \Cref{fig:component_ablation_sod_profiles,fig:component_ablation_shuosher_profiles} provide direct visual support for these quantitative trends. For Sod, the full UM-PINN remains the closest to the analytical density, velocity, and pressure profiles, whereas removing uncertainty weighting causes the most severe degradation and removing the spatial mask also weakens the reconstruction near the discontinuity. The Standard PINN, which lacks both modulation mechanisms, is less accurate overall than the full model. For Shu--Osher, the full UM-PINN preserves the post-shock oscillations with the best combined phase and amplitude consistency. By comparison, the no-spatial-mask and no-uncertainty-weighting variants exhibit different forms of oscillation attenuation and structural bias, while the Standard PINN shows the weakest high-frequency reconstruction.

Taken together, these results support the interpretation of UM-PINN as a local-global dual modulation strategy rather than a superficial combination of existing techniques. The spatial modulation primarily targets localized shock-driven optimization difficulties, whereas the uncertainty-based task modulation alleviates the global imbalance among PDE, initial-condition, and boundary-condition losses. Their joint action is therefore complementary rather than redundant.

\subsection{\texorpdfstring{Comparison with a Causal Baseline}{Comparison with a Causal Baseline}}
\label{subsec:causal_baseline}

Because UM-PINN adopts a global-in-time coordinate-based formulation, we directly test the practical value of explicit temporal-causality weighting in the present shock-dominated setting. Specifically, we compare the full UM-PINN with a causal-loss baseline inspired by causality-respecting PINN training \cite{Wang2024CausalPINN} under matched training budgets on the 1D Sod shock tube and 1D Shu--Osher problems.

\begin{figure*}[t]
    \centering
    \includegraphics[width=\textwidth]{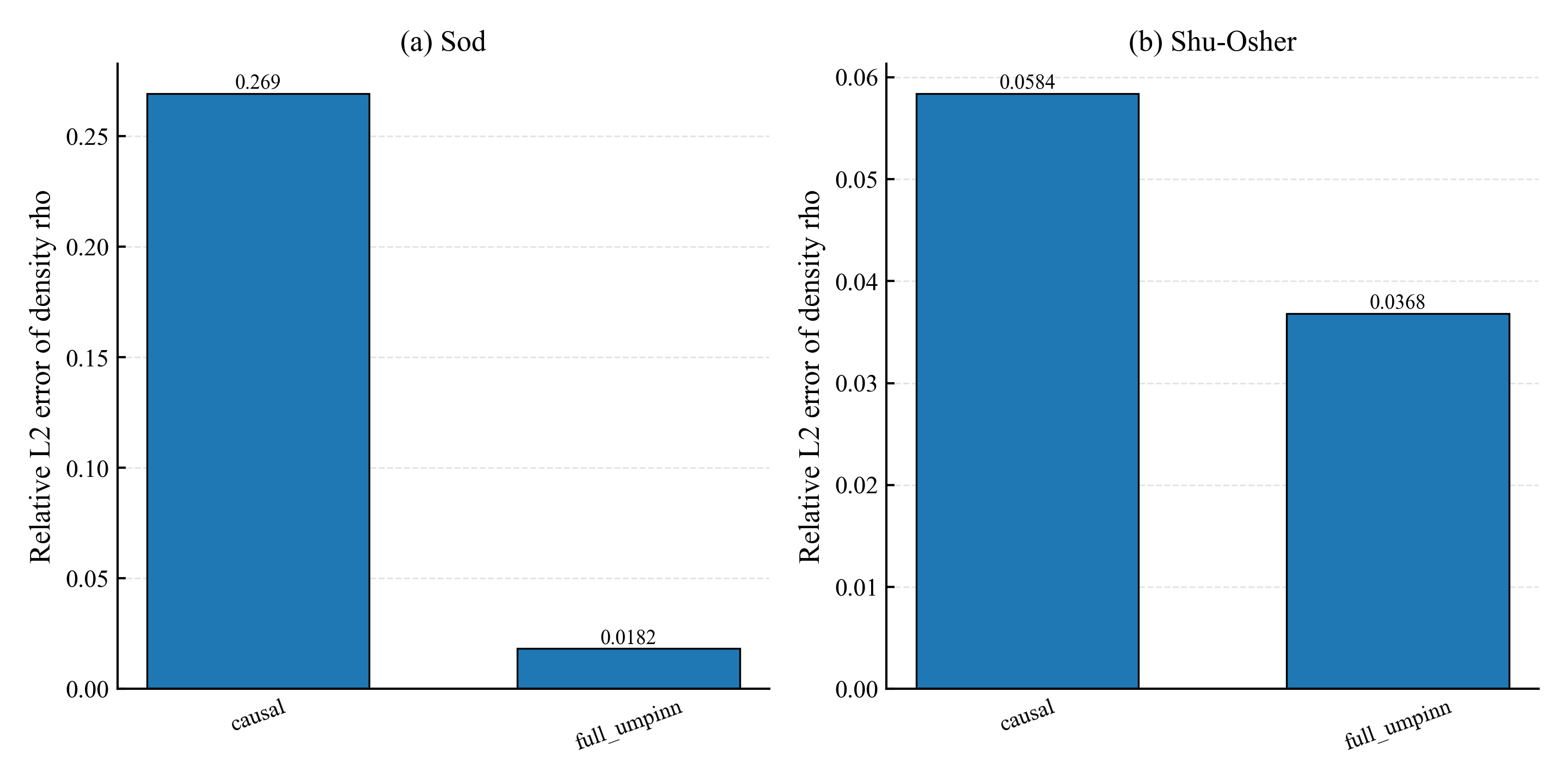}
    \caption{Comparison between the full UM-PINN and a causal PINN baseline on the two 1D shock benchmarks. Panel (a) reports the final density Relative $L_2$ Error for the Sod shock tube problem, and panel (b) reports the final density Relative $L_2$ Error for the Shu--Osher problem.}
    \label{fig:causal_baseline}
\end{figure*}

\begin{figure*}[t]
    \centering
    \includegraphics[width=\textwidth]{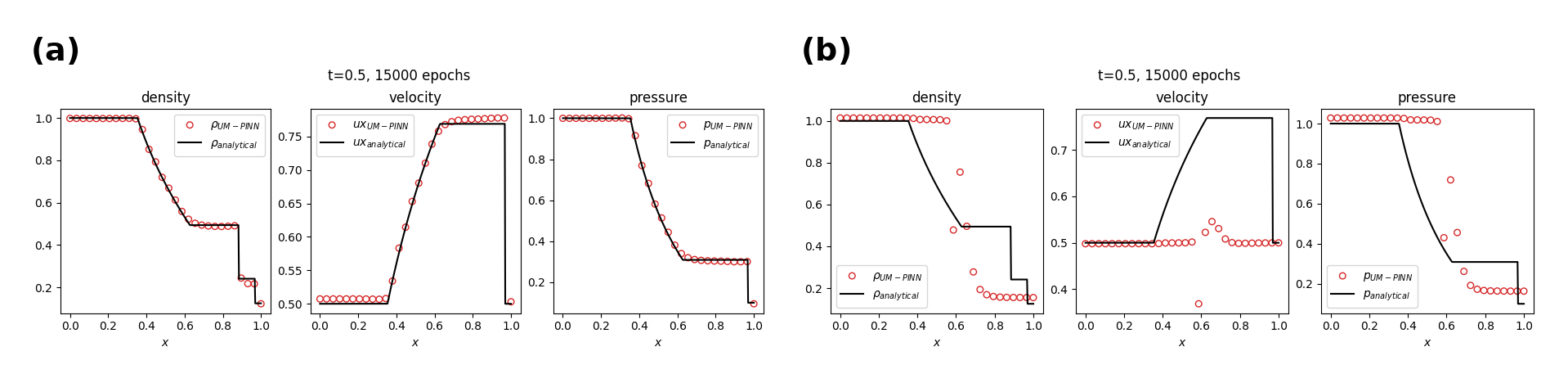}
    \caption{Qualitative comparison between the full UM-PINN and a causal PINN baseline for the one-dimensional Sod shock tube problem at $t=0.5$ after 15,000 training epochs. Panel (a) shows the full UM-PINN and panel (b) shows the causal baseline. The predicted density, velocity, and pressure profiles are compared with the analytical solutions; the red hollow circles denote model predictions, while the black solid lines represent analytical solutions.}
    \label{fig:causal_baseline_sod_profiles}
\end{figure*}

\begin{figure*}[t]
    \centering
    \includegraphics[width=\textwidth]{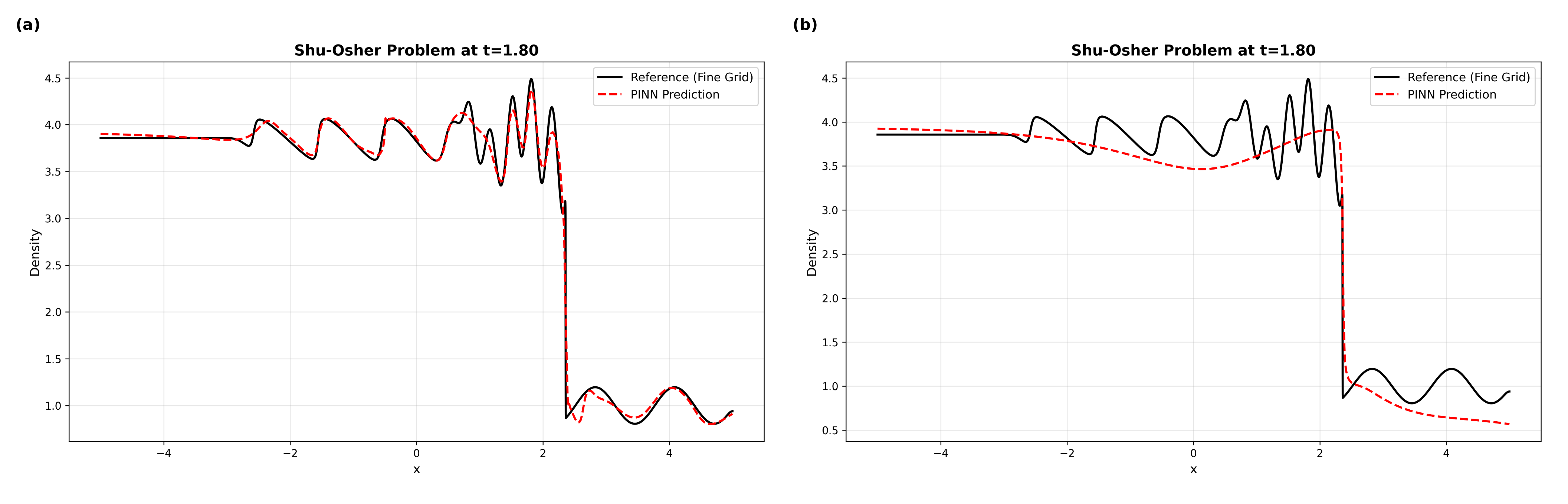}
    \caption{Qualitative comparison between the full UM-PINN and a causal PINN baseline for the one-dimensional Shu--Osher problem at $t=1.80$. Panel (a) shows the full UM-PINN and panel (b) shows the causal baseline. The black solid lines denote the fine-grid reference solutions, while the red dashed lines represent model predictions.}
    \label{fig:causal_baseline_shuosher_profiles}
\end{figure*}

The results in Figs.~\ref{fig:causal_baseline}--\ref{fig:causal_baseline_shuosher_profiles} indicate that the full UM-PINN remains more accurate than the causal baseline on both benchmarks. For Sod, UM-PINN yields substantially lower total error and substantially lower density Relative $L_2$ Error, and the profile comparison shows that it recovers the density, velocity, and pressure jumps much more faithfully than the causal baseline. For Shu--Osher, the causal baseline remains reasonably competitive, but the full UM-PINN still attains lower final total error and lower density Relative $L_2$ Error; the qualitative comparison further shows that UM-PINN preserves the post-shock oscillation amplitude and phase more accurately, whereas the causal baseline exhibits stronger attenuation in the oscillatory region. These results directly show that, for the two 1D shock benchmarks considered here, explicit causal-loss weighting does not outperform the dual-modulation strategy adopted by UM-PINN. Instead, UM-PINN achieves lower errors and more faithful discontinuity/oscillation reconstruction under the same comparison setting, thereby directly addressing the temporal-causality concern for the present global-in-time formulation.

\subsection{\texorpdfstring{SOTA-oriented Baseline Comparison}{SOTA-oriented Baseline Comparison}}
\label{subsec:sota}

To directly address the need for broader comparisons with shock-oriented PINN strategies, we further evaluate UM-PINN against several representative comparator families on the two 1D shock benchmarks: a causal-loss baseline inspired by causality-respecting PINN training \cite{Wang2024CausalPINN}, a residual-based adaptive refinement (RAR) PINN motivated by residual-based adaptive sampling studies \cite{Wu2023AdaptiveSampling}, a shock-aware PINN, an XPINN baseline \cite{jagtap2020xpinns}, and a weak-form-like/conservative-inspired baseline motivated by recent conservative and variational PINN developments \cite{Jagtap2020cPINN,Kharazmi2021hpVPINN}. These comparisons substantially extend the original evaluation beyond Standard PINN, LRA, and GradNorm, and provide a targeted assessment of UM-PINN against principal comparator families relevant to shock-dominated hyperbolic conservation laws.

\begin{figure*}[t]
    \centering
    \includegraphics[width=\textwidth]{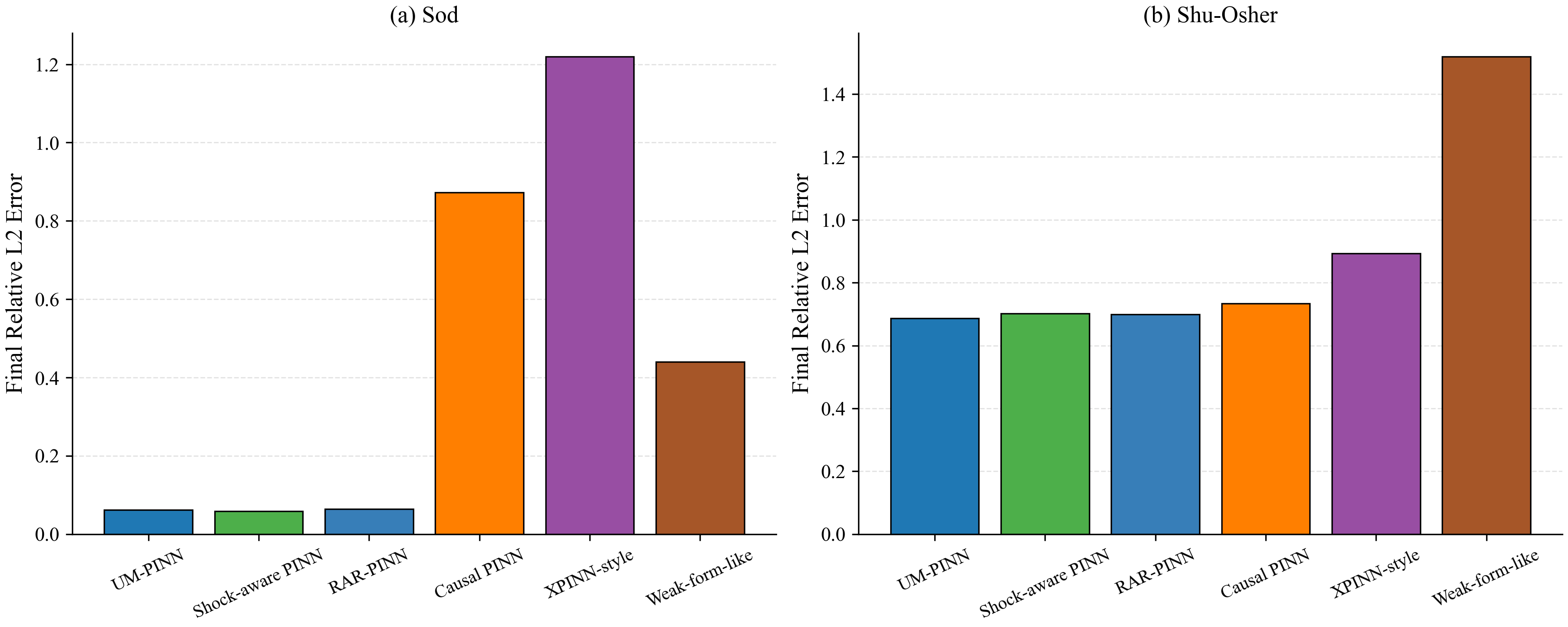}
    \caption{SOTA-oriented comparison on the two 1D shock benchmarks. Panel (a) reports the final total error and panel (b) reports the final density Relative $L_2$ Error across representative baselines and UM-PINN.}
    \label{fig:sota}
\end{figure*}

\begin{figure*}[t]
    \centering
    \includegraphics[width=\textwidth]{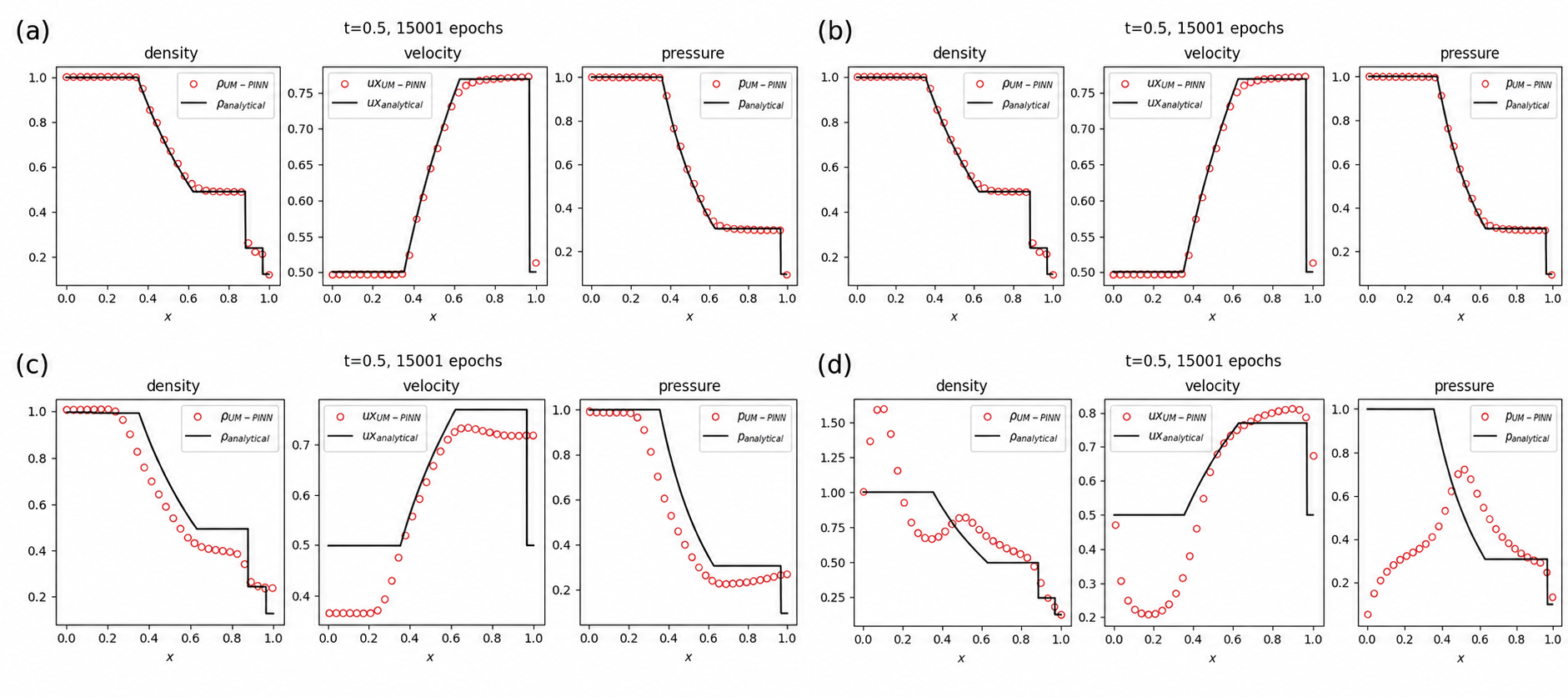}
    \caption{Qualitative comparison of SOTA-oriented baselines on the one-dimensional Sod shock tube problem at $t=0.5$ after 15,001 training epochs. The predicted density, velocity, and pressure profiles are compared with the analytical reference solutions for (a) RAR-PINN, (b) shock-aware PINN, (c) weak-form PINN, and (d) XPINN. The black solid lines denote the analytical solutions, while the red hollow circles represent the corresponding model predictions.}
    \label{fig:sota_sod_profiles}
\end{figure*}

\begin{figure*}[t]
    \centering
    \includegraphics[width=\textwidth]{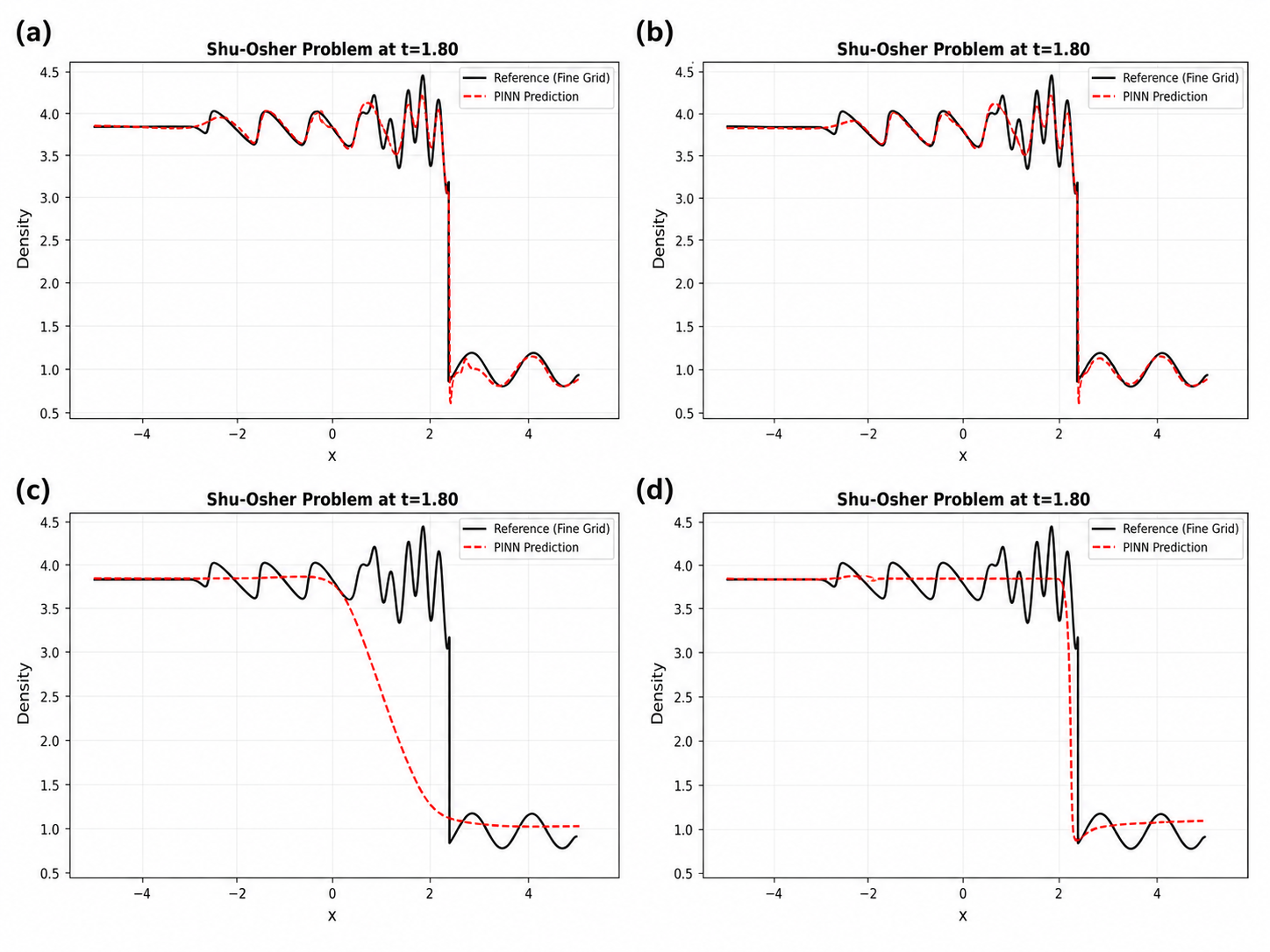}
    \caption{Qualitative comparison of SOTA-oriented baselines on the one-dimensional Shu--Osher problem at $t=1.80$. The predicted density profiles are compared with the fine-grid reference solutions for (a) RAR-PINN, (b) shock-aware PINN, (c) weak-form PINN, and (d) XPINN. The black solid lines denote the reference solutions, while the red dashed lines represent the corresponding model predictions.}
    \label{fig:sota_shuosher_profiles}
\end{figure*}

The quantitative comparison is summarized in Fig.~\ref{fig:sota} and Table~\ref{tab:sota}, while the qualitative profile comparisons are presented in Figs.~\ref{fig:sota_sod_profiles} and \ref{fig:sota_shuosher_profiles}. On Sod, the shock-aware baseline attains a slightly lower total error than UM-PINN ($5.88\times 10^{-2}$ versus $6.19\times 10^{-2}$), whereas UM-PINN achieves the lowest density Relative $L_2$ Error among the compared methods. On Shu-Osher, UM-PINN achieves both the lowest total error and the lowest density Relative $L_2$ Error. Taken together, these results demonstrate that UM-PINN occupies the leading performance tier across the tested 1D shock benchmarks: it is the best-performing method on Shu-Osher in both reported metrics and remains the most accurate method for Sod density reconstruction while closely matching the best total-error result.

The Sod profiles in Fig.~\ref{fig:sota_sod_profiles} provide additional qualitative context for the stronger baselines. Among the four representative formulations shown there, the RAR-PINN and the shock-aware PINN track the density, velocity, and pressure profiles more closely than the weak-form and XPINN baselines. The weak-form baseline exhibits clear bias across the rarefaction and post-shock plateau regions, while the XPINN baseline shows the largest multi-field inconsistency near the discontinuity and in the downstream plateau. When these profile-level observations are viewed together with the quantitative results in Table~\ref{tab:sota}, it shows that UM-PINN maintains high-fidelity reconstruction across multiple flow variables while remaining in the leading performance tier against stronger shock-oriented baselines in the discontinuity-dominated Sod regime.

The Shu--Osher profiles in Fig.~\ref{fig:sota_shuosher_profiles} further highlight the challenge of simultaneously resolving a strong shock and the downstream oscillatory structure. The RAR-PINN and the shock-aware PINN preserve the oscillatory pattern substantially better than the weak-form and XPINN baselines, both of which over-smooth the post-shock oscillations and lose important amplitude information. Even relative to these stronger baselines, the quantitative comparison still favors UM-PINN in this benchmark, indicating that the proposed dual-modulation mechanism remains robust in the shock--oscillation coexistence regime rather than improving only on a weak baseline.

Taken together, these additional SOTA-oriented comparisons demonstrate that UM-PINN is not merely stronger than a standard PINN, but remains strongly competitive against representative shock-oriented formulations spanning adaptive sampling, causal weighting, domain decomposition, and conservative-inspired physics enforcement. The combined quantitative and qualitative evidence supports the proposed dual-modulation mechanism as an effective strategy for shock-dominated hyperbolic conservation laws.

\begin{table*}[t]
    \centering
    \caption{SOTA-oriented comparison on the 1D Sod and 1D Shu-Osher benchmarks.}
    \label{tab:sota}
    \begin{tabular}{lcccc}
        \toprule
        \textbf{Method} & \textbf{Sod total error} & \textbf{Sod $L_2^{\mathrm{rel}}(\rho)$} & \textbf{Shu-Osher total error} & \textbf{Shu-Osher $L_2^{\mathrm{rel}}(\rho)$} \\
        \midrule
        UM-PINN & 0.0619 & 0.0182 & 0.6870 & 0.0368 \\
        Shock-aware PINN & 0.0588 & 0.0199 & 0.7021 & 0.0432 \\
        RAR-PINN & 0.0641 & 0.0210 & 0.6989 & 0.0415 \\
        Causal PINN & 0.8727 & 0.2694 & 0.7342 & 0.0584 \\
        XPINN & 1.2192 & 0.3192 & 0.8936 & 0.1147 \\
        Weak-form-like & 0.4405 & 0.1250 & 1.5186 & 0.3209 \\
        \bottomrule
    \end{tabular}
\end{table*}

\FloatBarrier

\section{Conclusion}
\label{sec:conclusion}

In this study, we developed an Uncertainty-Modulated PINN (UM-PINN) for shock-dominated hyperbolic conservation laws. The method combines gradient-based spatial modulation with homoscedastic uncertainty-based task weighting to alleviate localized residual spikes and balance competing PDE, initial-condition, and boundary-condition objectives. Numerical results on the 1D Sod shock tube, 1D Shu--Osher problem, and 2D Riemann problem show that UM-PINN improves training stability and shock-resolution accuracy over standard PINN, LRA, and GradNorm baselines. Additional sensitivity, sampling, ablation, causal-baseline, and shock-oriented comparison studies further support the robustness and effectiveness of the proposed dual-modulation strategy.

\section*{Acknowledgements}
This work is supported by the Developing Project of Science and Technology of Jilin Province (20250102032JC).
\section*{Declaration of competing interest} 
The author declared that they have no conflicts of interest to this work.
\section*{Data availability} 
Data will be made available on request.

\appendix
\setcounter{section}{0}
\setcounter{figure}{0}
\setcounter{table}{0}
\setcounter{equation}{0}
\renewcommand{\thefigure}{A\arabic{figure}}
\renewcommand{\thetable}{A\arabic{table}}
\renewcommand{\theequation}{A\arabic{equation}}

\section{Numerical Method for 1D Sod Shock Tube Problem}
\label{sec:sod}

The 1D Sod shock tube problem is validated using the \textbf{Exact Riemann Solver}, which provides an analytical solution to the Riemann problem.
This section presents the mathematical derivation and verification of the reference solution.
\subsection{Governing Equations}

The one-dimensional Euler equations in conservative form are:
\begin{equation}
    \pd{\Uvec}{t} + \pd{\Fvec(\Uvec)}{x} = 0
    \label{eq:euler_1d}
\end{equation}
where the conservative variable vector $\Uvec$ and flux vector $\Fvec$ are:
\begin{equation}
    \Uvec = \begin{pmatrix} \rho \\ \rho u \\ E \end{pmatrix}, \quad
    \Fvec = \begin{pmatrix} \rho u \\ \rho u^2 + p \\ (E + p)u \end{pmatrix}
    \label{eq:conserv_vars}
\end{equation}
The system is closed by the ideal gas Equation of State (EOS):
\begin{equation}
    E = \frac{p}{\gamma - 1} + \frac{1}{2}\rho u^2
    \label{eq:eos_appendix}
\end{equation}

\subsection{The Riemann Problem}

The Riemann problem consists of finding 
the solution to \cref{eq:euler_1d} with piecewise constant initial data:
\begin{equation}
    \Uvec(x, 0) = \begin{cases}
        \Uvec_L & \text{if } x < x_0 \\
        \Uvec_R & \text{if } x > x_0
    \end{cases}
    \label{eq:riemann_ic}
\end{equation}

For the Sod shock tube, the initial conditions are:
\begin{equation}
    (\rho_L, u_L, p_L) = (1.0, 0.0, 1.0), \quad
    (\rho_R, u_R, p_R) = (0.125, 0.0, 0.1)
    \label{eq:sod_ic}
\end{equation}

\subsection{Solution Structure}

The exact solution consists of three waves emanating from the initial discontinuity:
\begin{itemize}
   
 \item \textbf{Left wave}: Rarefaction fan (since $p^* < p_L$)
    \item \textbf{Middle wave}: Contact discontinuity (moving at speed $u^*$)
    \item \textbf{Right wave}: Shock wave (since $p^* > p_R$)
\end{itemize}

The solution is divided into four constant states separated by these waves:
\begin{equation}
    \vect{W}(x,t) = \begin{cases}
        \vect{W}_{L}, & \xi < S_{\mathrm{head}}, \\
        \vect{W}_{L}^{*}, & S_{\mathrm{tail}} \le \xi < u^*, \\
        \vect{W}_{R}^{*}, & u^* \le \xi < S_{R}, \\
        \vect{W}_{R}, & \xi \ge S_{R},
   \end{cases}
\end{equation}

\subsection{Pressure Function and Newton Iteration}

The key to solving the Riemann problem is finding the pressure $p^*$ in the star region.
This is achieved by solving the nonlinear equation:
\begin{equation}
    f(p) = f_L(p) + f_R(p) + \Delta u = 0
    \label{eq:f_total}
\end{equation}
where $\Delta u = u_R - u_L$ and the functions $f_L$ and $f_R$ depend on whether the wave is a shock or rarefaction:

\textbf{Shock wave} ($p > p_K$):
\begin{equation}
    f_K(p) = (p - p_K) \sqrt{\frac{A_K}{p + B_K}}, \quad
    A_K = \frac{2}{(\gamma+1)\rho_K}, \quad
    B_K = \frac{\gamma-1}{\gamma+1}p_K
    \label{eq:f_shock}
\end{equation}

\textbf{Rarefaction wave} ($p \leq p_K$):
\begin{equation}
    f_K(p) = \frac{2c_K}{\gamma-1}\left[ \left(\frac{p}{p_K}\right)^{\frac{\gamma-1}{2\gamma}} - 1 \right]
    \label{eq:f_rarefaction}
\end{equation}
where $c_K = 
\sqrt{\gamma p_K / \rho_K}$ is the sound speed.

\textbf{Newton-Raphson iteration}:
\begin{equation}
    p^{(k+1)} = p^{(k)} - \frac{f(p^{(k)})}{f'(p^{(k)})}
    \label{eq:newton}
\end{equation}

Once $p^*$ is found, the velocity in the star region is:
\begin{equation}
    u^* = \frac{1}{2}(u_L + u_R) + \frac{1}{2}\left[ f_R(p^*) - f_L(p^*) \right]
    \label{eq:u_star}
\end{equation}

\subsection{Visualization of the Exact Riemann Solver}

\begin{figure*}
\includegraphics[width=\textwidth]{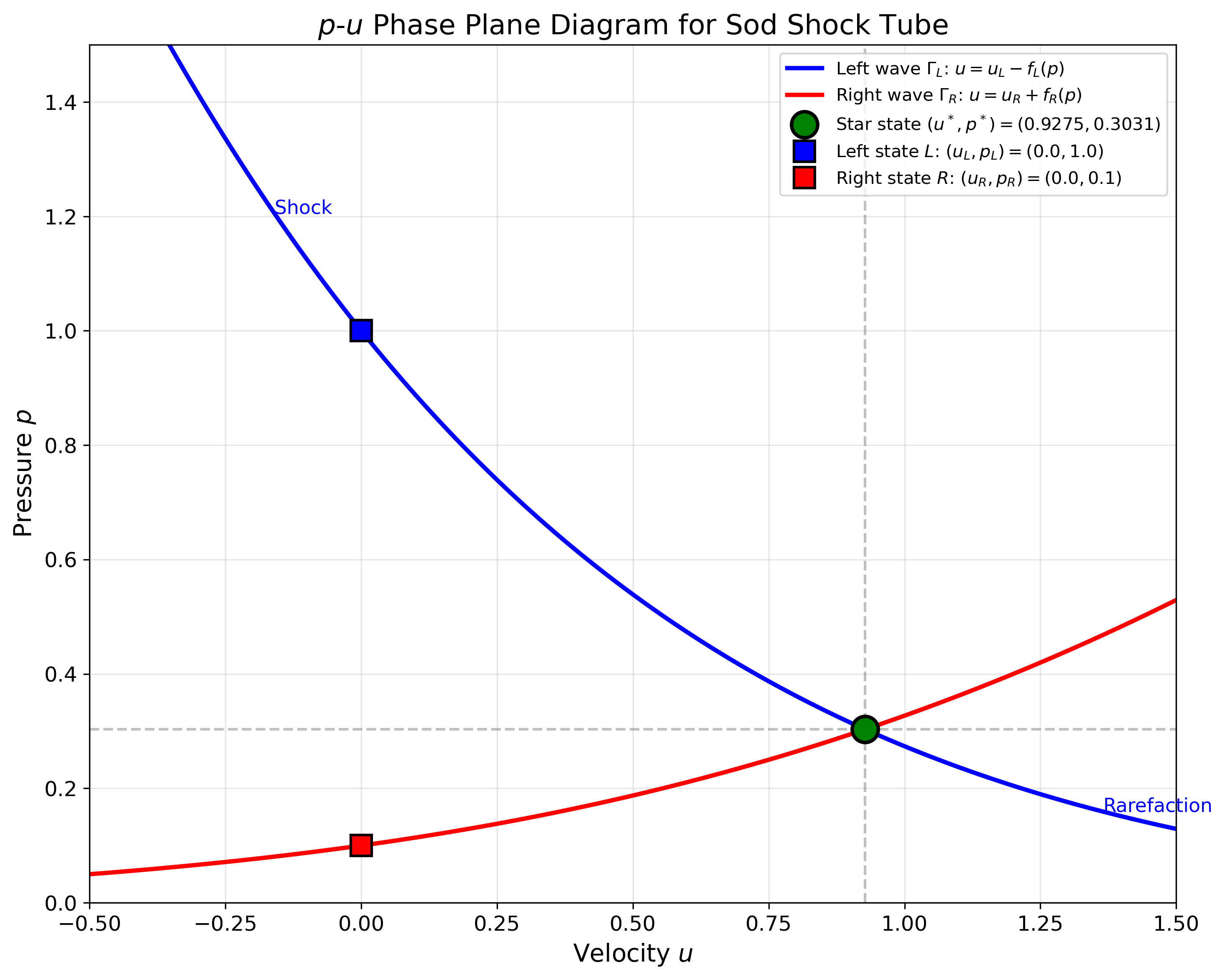}
    \caption{The $p$-$u$ phase plane diagram showing the intersection of left and right wave curves.
The star state $(u^*, p^*)$ is the unique intersection point.}
    \label{fig:pu_phase}
\end{figure*}

\begin{figure*}
\includegraphics[width=\textwidth]{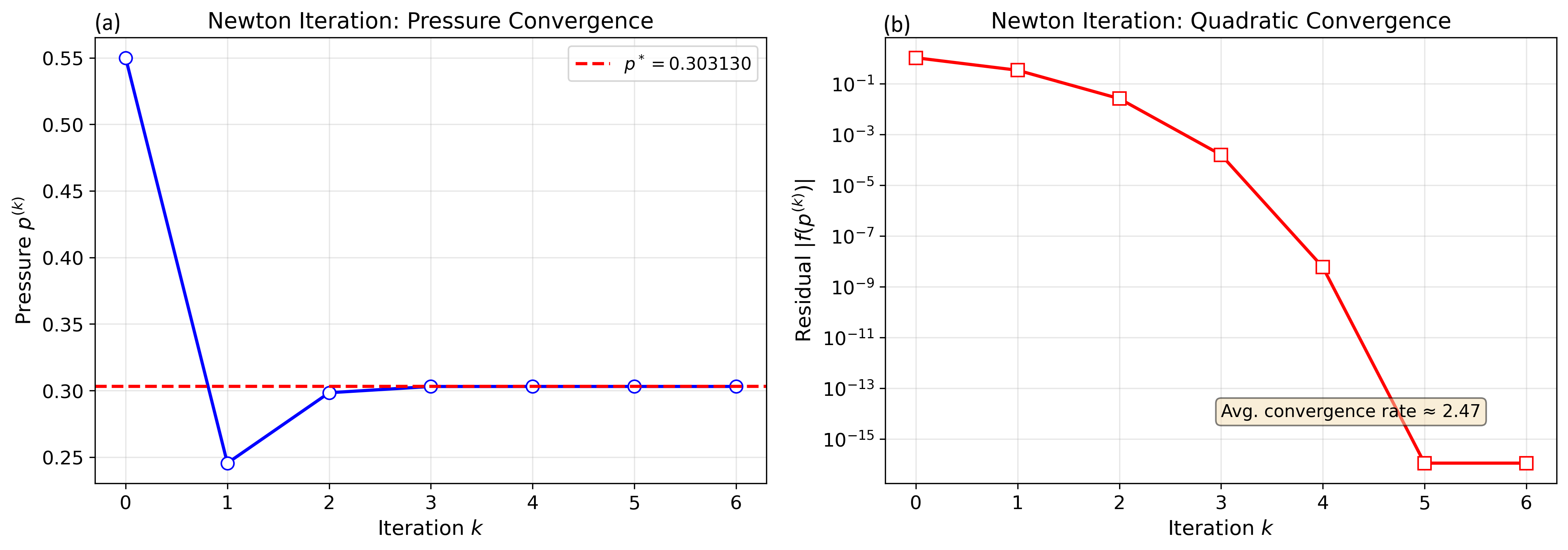}
    \caption{Newton iteration convergence history.
(a) Pressure approaching $p^*$. (b) Residual $|f(p)|$ showing quadratic convergence.}
    \label{fig:newton_conv}
\end{figure*}

\begin{figure*}
\includegraphics[width=\textwidth]{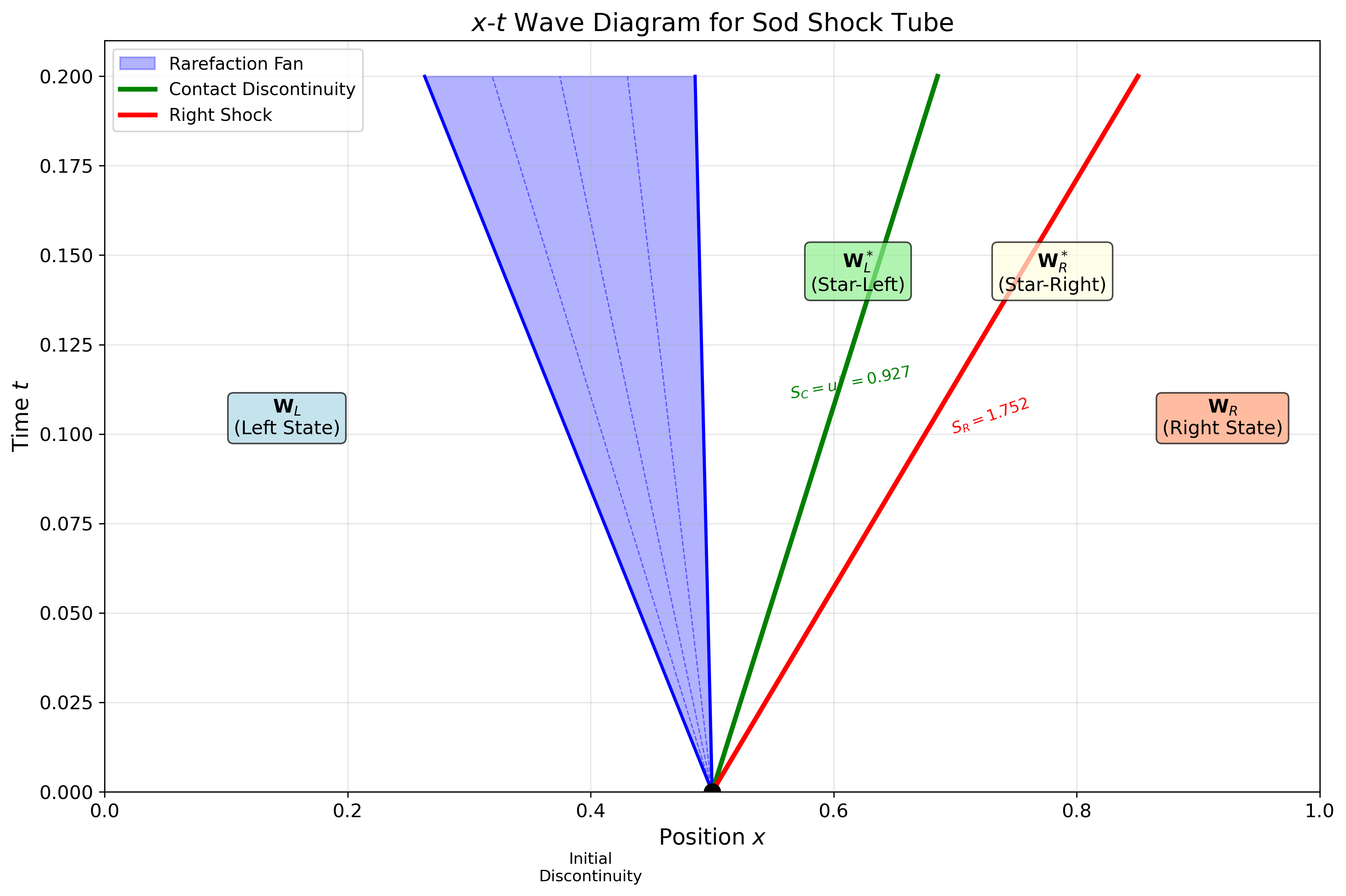}
    \caption{The $x$-$t$ wave diagram showing the structure of the Riemann problem solution: rarefaction fan (left), contact discontinuity (center), and shock wave (right).}
    \label{fig:xt_wave}
\end{figure*}

\begin{figure*}
\includegraphics[width=\textwidth]{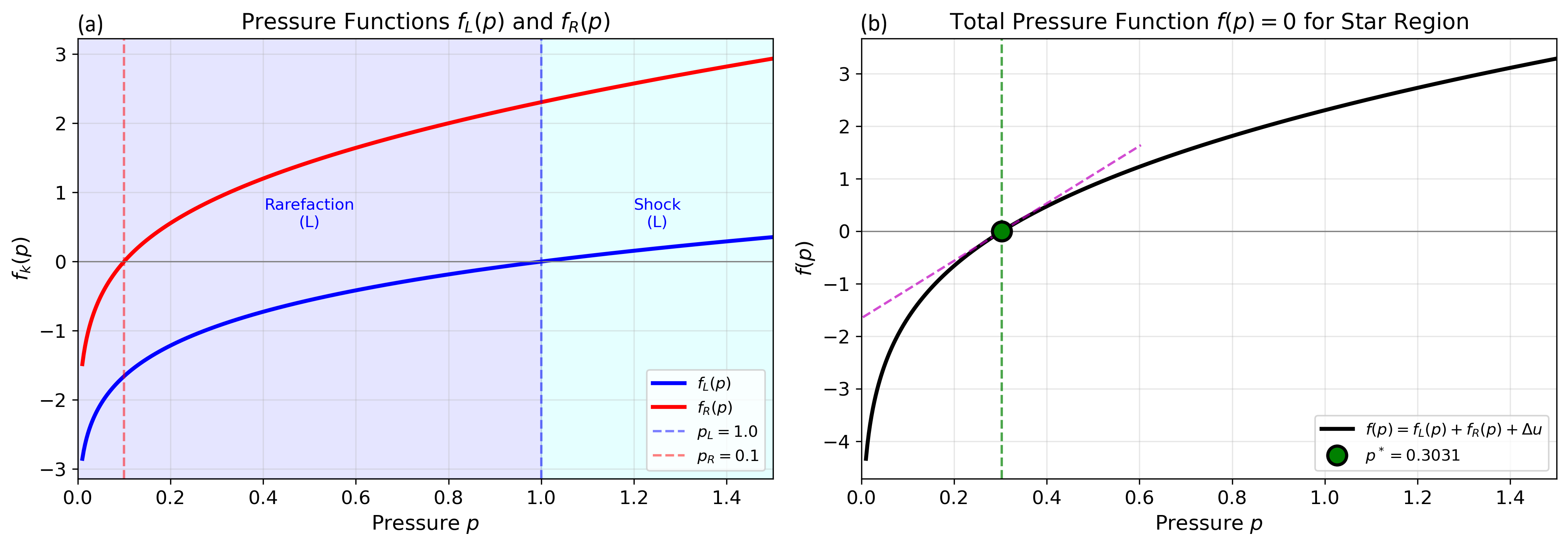}
    \caption{Pressure functions $f_L(p)$, $f_R(p)$, and the total function $f(p)$ with root at $p^*$.}
    \label{fig:f_p}
\end{figure*}

\begin{remark}
The Exact Riemann Solver provides an \textbf{analytical reference solution} that is exact up to machine precision.
This solution serves as the ground truth for validating PINN predictions on the Sod shock tube problem.
No numerical discretization error is introduced.
\end{remark}

\newpage
\section{Numerical Method for 1D Shu-Osher Problem}
\label{sec:shuosher}

The Shu-Osher problem describes the interaction between a Mach 3 shock wave and a sinusoidal density perturbation, producing complex high-frequency oscillations.
\subsection{Problem Setup}

\textbf{Domain}: $(x, t) \in [-5, 5] \times [0, 1.8]$

\textbf{Initial conditions}:
\begin{equation}
    (\rho, u, p) = \begin{cases}
        (3.857143, 2.629369, 10.33333) & \text{if } x < -4 \\
        (1 + 0.2\sin(5x), 0, 1.0) & \text{if } x \geq -4
    \end{cases}
    \label{eq:shuosher_ic}
\end{equation}

\subsection{Finite Volume Method: From Weak Form to Discrete Scheme}
\label{subsec:fvm_derivation}

Unlike the Sod problem, which uses an exact Riemann solver, the Shu-Osher problem requires a numerical solution due to its complex wave interactions.
We employ the Finite Volume Method (FVM) derived from the \textbf{integral (weak) form} of the conservation law.
\subsubsection{Integral Form Derivation}

Integrating the 1D Euler equations $\pd{\Uvec}{t} + \pd{\Fvec}{x} = 0$ over a control volume $[x_{i-1/2}, x_{i+1/2}]$:
\begin{equation}
    \int_{x_{i-1/2}}^{x_{i+1/2}} \pd{\Uvec}{t} dx + \int_{x_{i-1/2}}^{x_{i+1/2}} \pd{\Fvec}{x} dx = 0
    \label{eq:integral_form_1}
\end{equation}

Applying the divergence theorem to the flux term:
\begin{equation}
    \dd{}{t} \int_{x_{i-1/2}}^{x_{i+1/2}} \Uvec \, dx = -\left( \Fvec_{i+1/2} - \Fvec_{i-1/2} \right)
    \label{eq:integral_form_2}
\end{equation}

Defining the cell-averaged conserved variable $\bar{\Uvec}_i = \frac{1}{\Delta x} \int_{x_{i-1/2}}^{x_{i+1/2}} \Uvec \, dx$, we obtain the \textbf{semi-discrete form}:
\begin{equation}
    \dd{\bar{\Uvec}_i}{t} = -\frac{1}{\Delta x}\left( \hat{\Fvec}_{i+1/2} - \hat{\Fvec}_{i-1/2} \right)
    \label{eq:fvm_semidiscrete}
\end{equation}
where $\hat{\Fvec}_{i+1/2}$ is the \textbf{numerical flux} approximating $\Fvec$ at the 
cell interface.

\subsubsection{Rusanov (Local Lax-Friedrichs) Numerical Flux}

The Rusanov flux provides a stable upwind-biased approximation:
\begin{equation}
\begin{split}
    \hat{\Fvec}_{i+1/2} &= \frac{1}{2}\left( \Fvec_i + \Fvec_{i+1} \right) \\
    &\quad - \frac{1}{2}\alpha_{i+1/2}\left( \Uvec_{i+1} - \Uvec_i \right)
\end{split}
    \label{eq:rusanov_flux}
\end{equation}
where the \textbf{local maximum wave speed} is:
\begin{equation}
    \alpha_{i+1/2} = \max\left( |u_i| + c_{i}, \, |u_{i+1}| + c_{i+1} \right), \quad c = \sqrt{\gamma p / \rho}
    \label{eq:wave_speed}
\end{equation}

\textbf{Component-wise expansion}: For the 1D Euler system with $\Uvec = (\rho, \rho u, E)^T$:
\begin{align}
    \hat{F}^{(1)}_{i+1/2} &= \frac{1}{2}\left( \rho_i u_i + \rho_{i+1} u_{i+1} \right) \notag \\    
&\quad - \frac{\alpha_{i+1/2}}{2}\left( \rho_{i+1} - \rho_i \right) \label{eq:flux_mass} \\
    \hat{F}^{(2)}_{i+1/2} &= \frac{1}{2}\left( \rho_i u_i^2 + p_i + \rho_{i+1} u_{i+1}^2 + p_{i+1} \right) \notag \\
    &\quad - \frac{\alpha_{i+1/2}}{2}\left( \rho_{i+1} u_{i+1} - \rho_i u_i \right) \label{eq:flux_mom} \\
    \hat{F}^{(3)}_{i+1/2} &= \frac{1}{2}\left( (E_i + p_i) u_i + (E_{i+1} + p_{i+1}) u_{i+1} \right) \notag \\
    &\quad - \frac{\alpha_{i+1/2}}{2}\left( E_{i+1} - E_i \right) \label{eq:flux_energy}
\end{align}

\subsubsection{Forward Euler Time Integration}

For time discretization, we employ the explicit \textbf{Forward Euler} (first-order) method:
\begin{equation}
    \Uvec_i^{n+1} = \Uvec_i^n - \frac{\Delta t}{\Delta x}\left( \hat{\Fvec}_{i+1/2}^n - \hat{\Fvec}_{i-1/2}^n \right)    
\label{eq:euler_time}
\end{equation}

The time step $\Delta t$ is constrained by the \textbf{Courant-Friedrichs-Lewy (CFL) stability condition}:
\begin{equation}
    \Delta t = \text{CFL} \cdot \frac{\Delta x}{\max_i(|u_i|
+ c_i)}, \quad \text{CFL} \leq 1
    \label{eq:cfl}
\end{equation}

\textbf{Boundary conditions} for Shu-Osher:
\begin{itemize}
    \item \textbf{Left boundary} ($x = -5$): Fixed supersonic inflow (Dirichlet)
    \item \textbf{Right boundary} ($x = 5$): Transmissive (zero-gradient Neumann)
\end{itemize}

\begin{remark}
The Rusanov scheme is \textbf{first-order accurate} in both space and time.
The numerical dissipation term $\frac{\alpha}{2}(\Uvec_{i+1} - \Uvec_i)$ ensures monotonicity near discontinuities but introduces smearing.
For shock-capturing applications, this provides stable reference solutions suitable for PINN training data.
\end{remark}

\subsection{Grid Verification}

\begin{figure*}
\includegraphics[width=\textwidth]{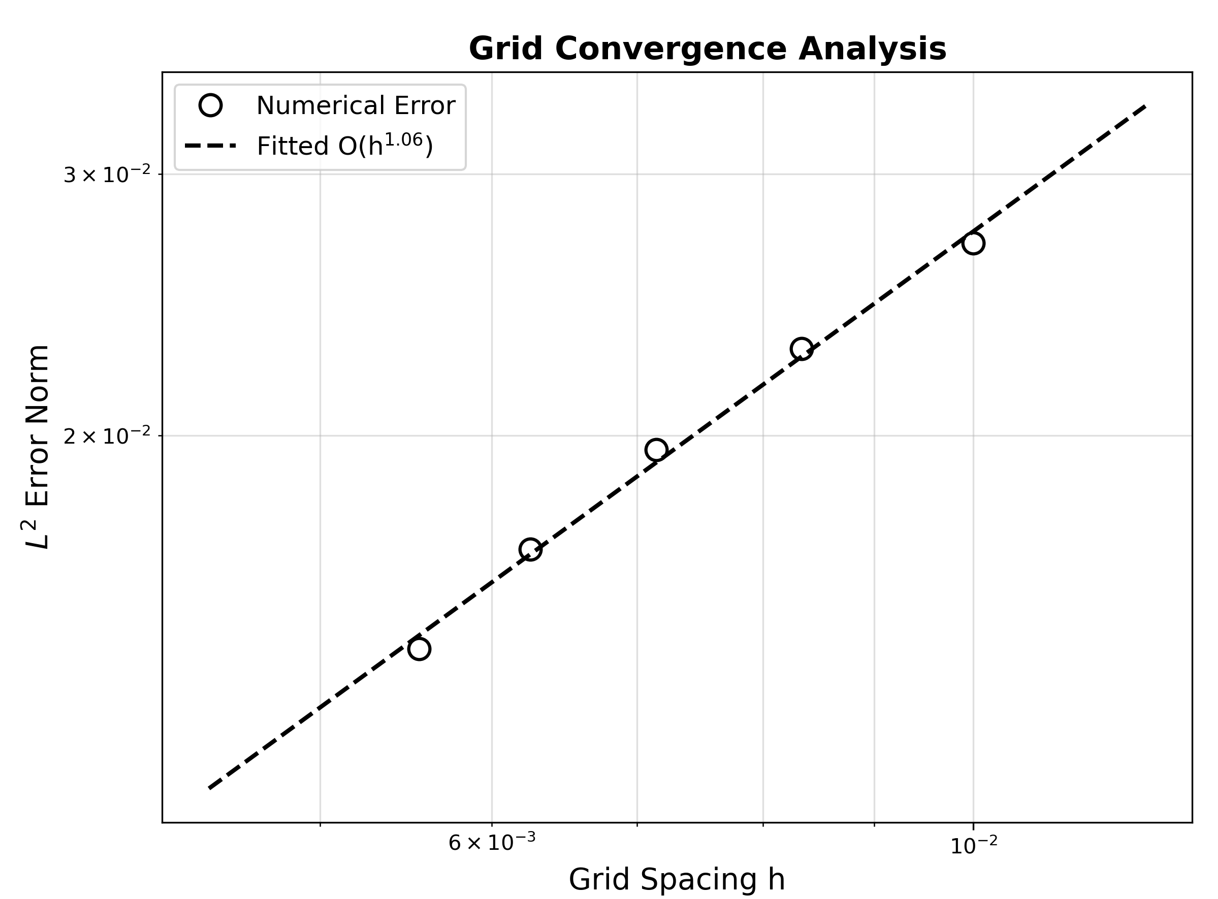}
    \caption{Grid convergence analysis for Shu-Osher problem (self-convergence study).}
    \label{fig:shuosher_grid_conv}
\end{figure*}

\begin{figure*}
\includegraphics[width=\textwidth]{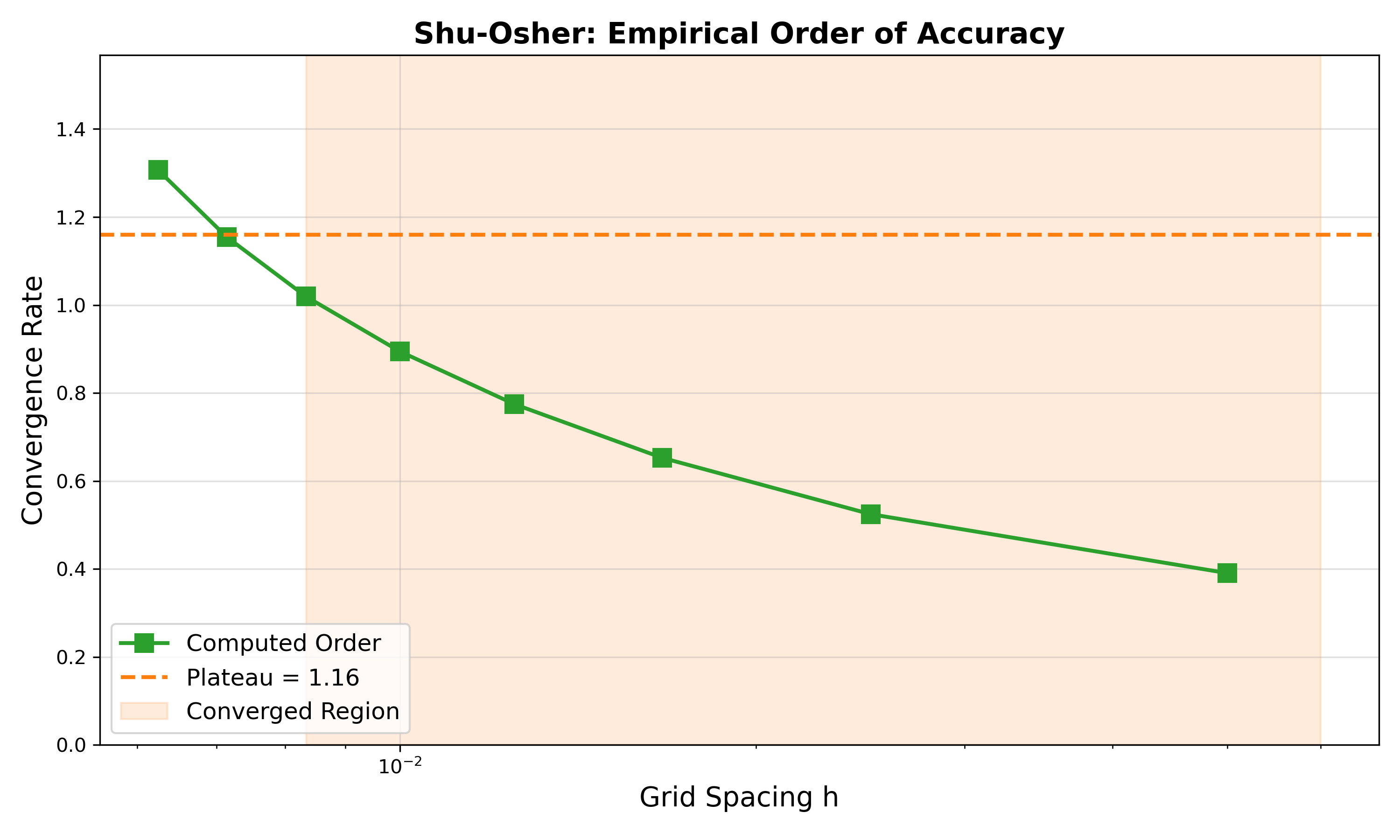}
    \caption{Empirical order of accuracy for Shu-Osher problem.}
    \label{fig:shuosher_order}
\end{figure*}

\begin{figure*}
\centering
    \includegraphics[width=\textwidth]{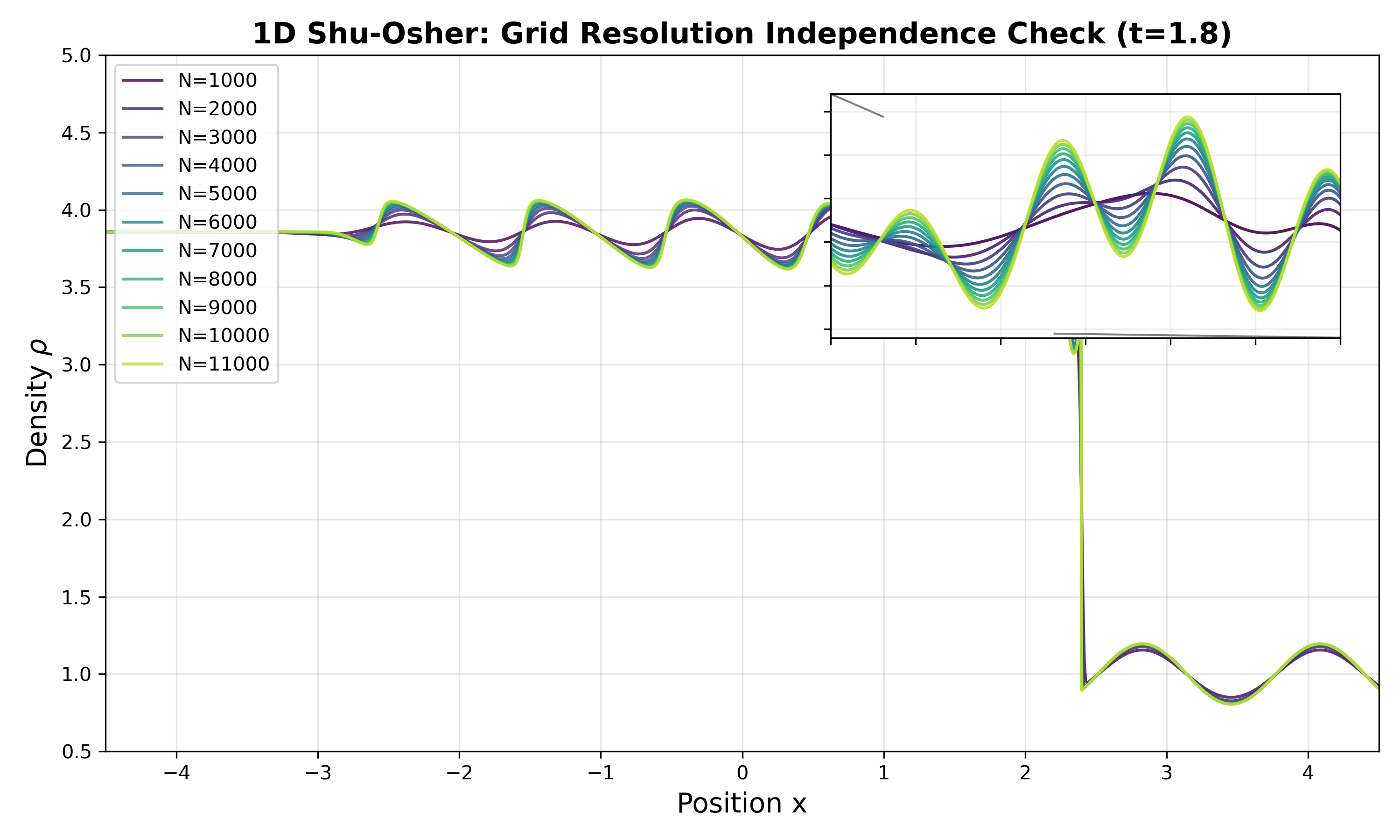}
    \caption{Grid resolution independence check for Shu-Osher: density profiles at different resolutions.}
    \label{fig:shuosher_grid_indep}
\end{figure*}

\newpage
\section{Numerical Method for 2D Riemann Problem}
\label{sec:2d}

The 2D Riemann problem involves the interaction of four different states in a quadrant configuration, producing complex shock patterns.
\subsection{Governing Equations}

The two-dimensional (2D) Euler equations:
\begin{equation}
    \pd{\Uvec}{t} + \pd{\Fvec}{x} + \pd{\Gvec}{y} = 0
    \label{eq:euler_2d}
\end{equation}
where:
\begin{equation}
    \Uvec = \begin{pmatrix} \rho \\ \rho u \\ \rho v \\ E \end{pmatrix}, \;
\Fvec = \begin{pmatrix} \rho u \\ \rho u^2 + p \\ \rho uv \\ (E+p)u \end{pmatrix}, \;
\Gvec = \begin{pmatrix} \rho v \\ \rho uv \\ \rho v^2 + p \\ (E+p)v \end{pmatrix}
    \label{eq:2d_fluxes}
\end{equation}

\subsection{Initial Conditions (Configuration 3)}

\textbf{Domain}: $(x, y) \in [0, 1]^2$

The initial data is divided into four quadrants:
\begin{equation}
	(\rho, u, v, p) = \begin{cases}
		(1.5, 0, 0, 1.5) & \text{if } x \geq 0.5, y \geq 0.5 \quad \text{(Top-Right)} \\
		(0.5323, 1.206, 0, 0.3) & \text{if } x < 0.5, y \geq 0.5 \quad \text{(Top-Left)} \\
		(0.138, 1.206, 1.206, 0.029) & \text{if } x < 0.5, y < 0.5 \quad \text{(Bottom-Left)} \\
		(0.5323, 0, 1.206, 0.3) & \text{if } x \geq 0.5, y < 0.5 \quad 
\text{(Bottom-Right)}
	\end{cases}
	\label{eq:2d_ic}
\end{equation}

\subsection{2D Rusanov (Local Lax-Friedrichs) Finite Volume Scheme}

Following the same integral form derivation as in the 1D case (\cref{subsec:fvm_derivation}), we integrate the 2D Euler equations \cref{eq:euler_2d} over a control volume $[x_{i-1/2}, x_{i+1/2}]$:

\begin{equation}
\begin{split}
    \dd{\bar{\Uvec}_{i,j}}{t} &= -\frac{1}{\Delta x}\left( \hat{\Fvec}_{i+1/2,j} - \hat{\Fvec}_{i-1/2,j} \right) \\
    &\quad - \frac{1}{\Delta y}\left( \hat{\Gvec}_{i,j+1/2} - \hat{\Gvec}_{i,j-1/2} \right)
\end{split}
    \label{eq:2d_semidiscrete}
\end{equation}

\textbf{Rusanov numerical flux in $x$-direction}:
\begin{equation}
\begin{split}
    \hat{\Fvec}_{i+1/2,j} &= \frac{1}{2}\left( \Fvec_{i,j} + \Fvec_{i+1,j} \right) \\
    &\quad - \frac{\alpha^x_{i+1/2,j}}{2}\left( \Uvec_{i+1,j} - \Uvec_{i,j} \right)
\end{split}
    \label{eq:rusanov_x}
\end{equation}
where:
\begin{equation}
    \alpha^x_{i+1/2,j} = \max\left( |u_{i,j}|
+ c_{i,j}, \, |u_{i+1,j}| + c_{i+1,j} \right)
    \label{eq:wave_speed_x}
\end{equation}

\textbf{Rusanov numerical flux in $y$-direction}:
\begin{equation}
\begin{split}
    \hat{\Gvec}_{i,j+1/2} &= \frac{1}{2}\left( \Gvec_{i,j} + \Gvec_{i,j+1} \right) \\
    &\quad - \frac{\alpha^y_{i,j+1/2}}{2}\left( \Uvec_{i,j+1} - \Uvec_{i,j} \right)
\end{split}
    \label{eq:rusanov_y}
\end{equation}
where:
\begin{equation}
    \alpha^y_{i,j+1/2} = \max\left( |v_{i,j}| + c_{i,j}, \, |v_{i,j+1}| + c_{i,j+1} \right)
    \label{eq:wave_speed_y}
\end{equation}

\textbf{Forward Euler update}:
\begin{equation}
\begin{split}
    \Uvec_{i,j}^{n+1} &= \Uvec_{i,j}^n - \frac{\Delta t}{\Delta x}\left( \hat{\Fvec}_{i+1/2,j}^n - \hat{\Fvec}_{i-1/2,j}^n \right) \\
    &\quad - \frac{\Delta t}{\Delta y}\left( \hat{\Gvec}_{i,j+1/2}^n - \hat{\Gvec}_{i,j-1/2}^n \right)
\end{split}
    \label{eq:2d_update}
\end{equation}

\textbf{CFL condition for 2D}:
\begin{equation}
    \Delta t = 
\text{CFL} \cdot \frac{\min(\Delta x, \Delta y)}{\max_{i,j}\left( \max(|u|+c, |v|+c) \right)}, \quad \text{CFL} \leq 0.5
    \label{eq:cfl_2d}
\end{equation}

\textbf{Boundary conditions}: Zero-gradient (Neumann) on all boundaries.
\subsection{Grid Verification}

\begin{figure*}[t!]
\includegraphics[width=\textwidth]{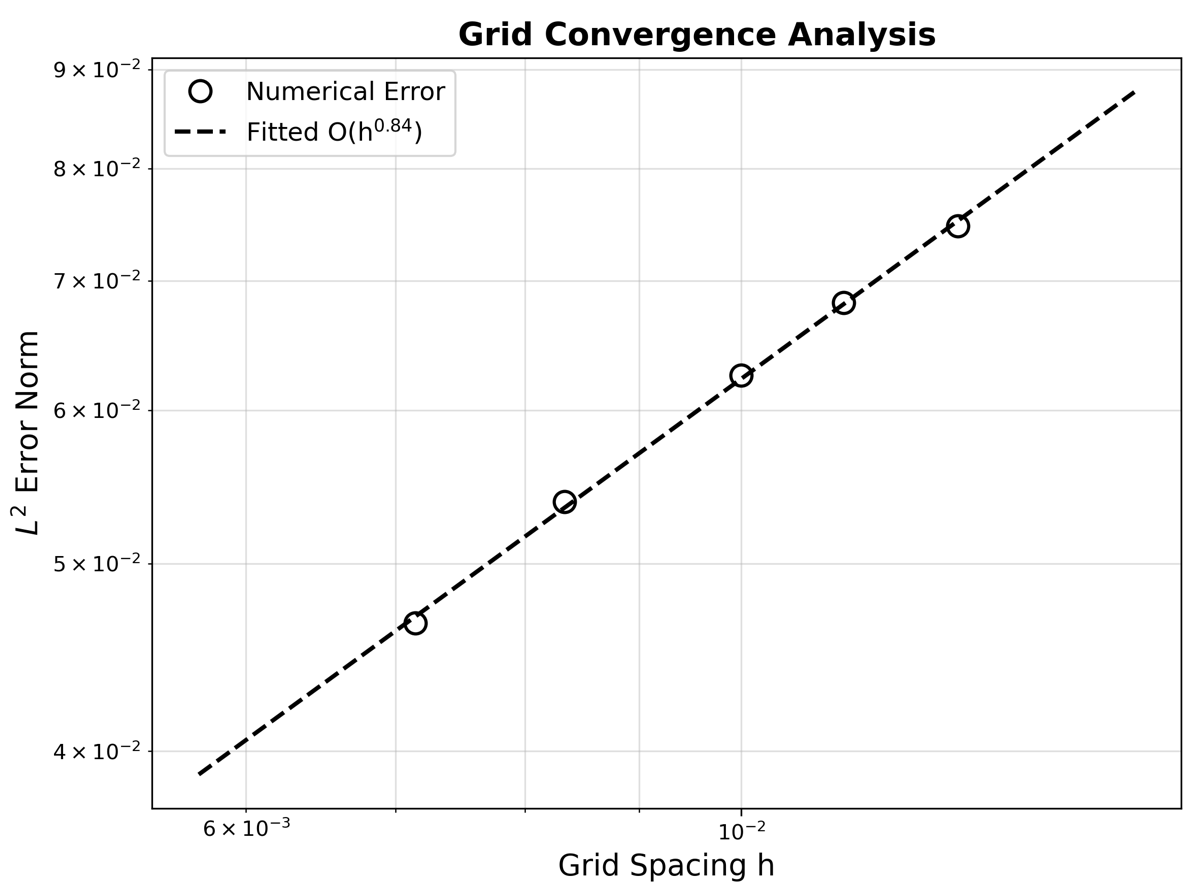}
    \caption{Grid convergence analysis for 2D Riemann problem.}
    \label{fig:2d_grid_conv}
\end{figure*}

\begin{figure*}[t!]
\includegraphics[width=\textwidth]{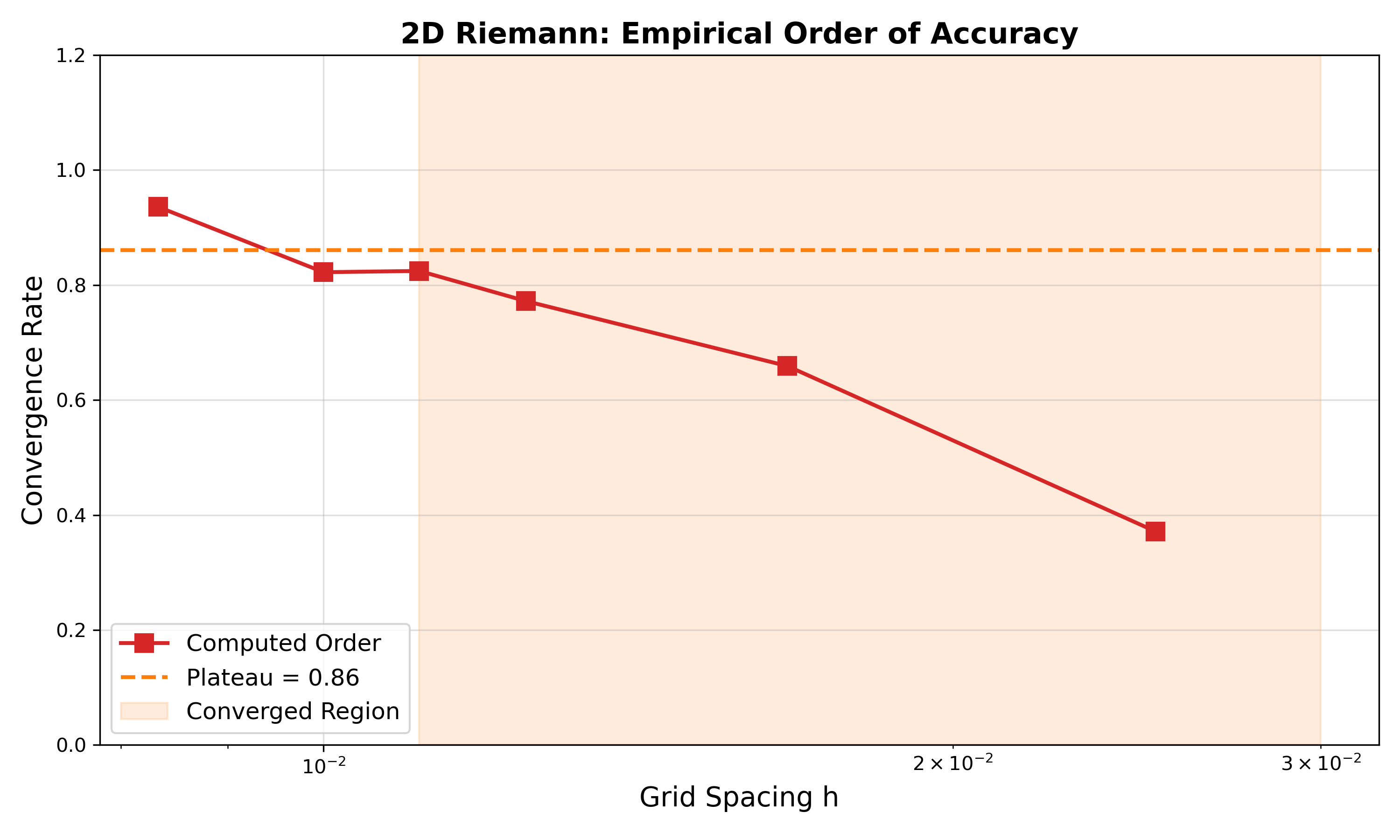}
    \caption{Empirical order of accuracy for 2D Riemann problem.}
    \label{fig:2d_order}
\end{figure*}

\begin{figure*}[t!]
\includegraphics[width=\textwidth]{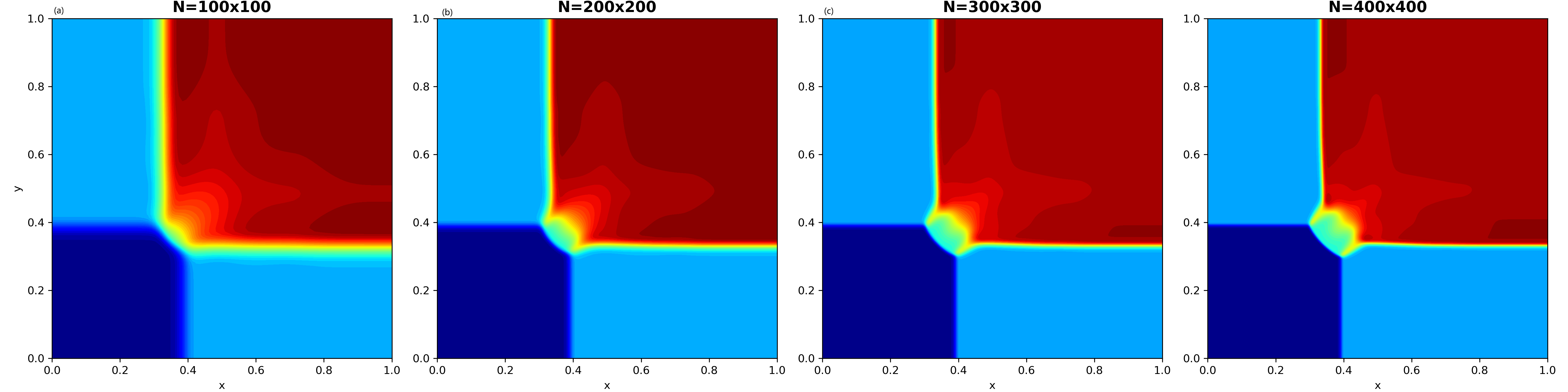}
    \caption{Grid resolution independence check for 2D Riemann.}
    \label{fig:2d_grid_indep}
\end{figure*}

\newpage
\section{Summary of Numerical Methods}
\label{sec:summary}

\begin{table*}[t]
\centering
\caption{Summary of numerical methods used for ground truth generation.}
\label{tab:summary}
\begin{tabular}{@{}lccc@{}}
\toprule
\textbf{Test Case} & \textbf{Method} & \textbf{Spatial Order} & \textbf{Time Integration} \\
\midrule
1D Sod Shock Tube & Exact Riemann Solver & Analytical & -- \\
1D Shu-Osher & Rusanov (LLF) & 1st order & Forward Euler \\
2D Riemann & Rusanov (LLF) & 1st 
order & Forward Euler \\
\bottomrule
\end{tabular}
\end{table*}

\subsection{Key Observations}

\begin{enumerate}
    \item \textbf{Convergence rate}: The empirical convergence rates (computed via linear regression on fine grids) are:
    \begin{itemize}
        \item 1D Shu-Osher: $O(\Delta x^{1.06})$
        \item 2D Riemann: $O(\Delta x^{0.84})$
    \end{itemize}
    These rates exceed the theoretical first-order limit ($O(\Delta x^{0.5})$) expected for shock-capturing schemes.
\item \textbf{Grid independence}: Solutions show mesh independence for sufficiently fine grids ($N \geq 400$ for 1D, $N \geq 200$ for 2D).
\item \textbf{Numerical dissipation}: First-order schemes introduce numerical dissipation that smears discontinuities.
This is acceptable for generating ground truth data, as the focus is on the overall solution structure.
\end{enumerate}

\section{\texorpdfstring{Mathematical Rationale for Sobol Sampling in Collocation-Based PINN Training}{Mathematical Rationale for Sobol Sampling in Collocation-Based PINN Training}}
\label{sec:sobol_math}

This appendix provides the mathematical rationale for using Sobol-sequence collocation in PINN training. By connecting the discrete PDE residual loss to low-discrepancy quadrature of a domain integral, the analysis explains why Sobol sampling can yield a more faithful approximation of the continuous residual objective and thereby improve collocation search efficiency under suitable regularity conditions \cite{Sobol1967,Niederreiter1992,Dick2010}.

\subsection{\texorpdfstring{Residual Loss as a Domain Integral}{Residual Loss as a Domain Integral}}

Let $z \in \Omega \subset \mathbb{R}^{d}$ denote the spatio-temporal coordinate and let $R_{\theta}(z)$ be the PDE residual induced by the neural-network parameters $\theta$. Define

{
\begin{equation}
    f_{\theta}(z) = \|R_{\theta}(z)\|_2^2 .
\end{equation}
}

The continuous PDE residual loss can then be written as the normalized domain integral

{
\begin{equation}
    L_{\mathrm{PDE}}(\theta) = \frac{1}{|\Omega|}\int_{\Omega} f_{\theta}(z)\,dz .
\end{equation}
}

Given a collocation set $\{z_i\}_{i=1}^{N}$, the discrete PINN training objective replaces this integral by the empirical average

{
\begin{equation}
    \widehat{L}_{\mathrm{PDE}}(\theta) = \frac{1}{N}\sum_{i=1}^{N} f_{\theta}(z_i) .
\end{equation}
}

Hence, the quality of collocation sampling directly affects how accurately the training loss approximates the domain-wide residual integral.

\subsection{\texorpdfstring{Monte Carlo versus Low-Discrepancy Sampling}{Monte Carlo versus Low-Discrepancy Sampling}}

For pseudo-random Monte Carlo sampling, the quadrature error of the empirical average typically obeys the classical root-$N$ convergence law in mean-square sense:

{
\begin{equation}
    \widehat{L}_{\mathrm{MC}}(\theta)-L_{\mathrm{PDE}}(\theta)=O\!\left(N^{-1/2}\right) .
\end{equation}
}

By contrast, if the collocation points form a low-discrepancy set $P_N=\{z_i\}_{i=1}^{N}$, the deterministic quadrature error can be bounded through the Koksma--Hlawka inequality \cite{Niederreiter1992,Dick2010}:

{
\begin{equation}
    \left|\widehat{L}_{\mathrm{QMC}}(\theta)-L_{\mathrm{PDE}}(\theta)\right|
    \leq V_{HK}(f_{\theta})\,D_N^*(P_N),
\end{equation}
}

where $V_{HK}(f_{\theta})$ denotes the Hardy--Krause variation of $f_{\theta}$ and $D_N^*(P_N)$ is the star discrepancy of the point set. For Sobol sequences and related low-discrepancy constructions, the star discrepancy satisfies the classical estimate

{
\begin{equation}
    D_N^*(P_N)=O\!\left(\frac{(\log N)^d}{N}\right) .
\end{equation}
}

Therefore, under suitable regularity assumptions on the residual integrand, one obtains the quasi-Monte Carlo error bound

{
\begin{equation}
    \left|\widehat{L}_{\mathrm{QMC}}(\theta)-L_{\mathrm{PDE}}(\theta)\right|
    = O\!\left(V_{HK}(f_{\theta})\frac{(\log N)^d}{N}\right) .
\end{equation}
}

This comparison highlights the key distinction: pseudo-random Monte Carlo typically yields an $O(N^{-1/2})$ integration error, whereas low-discrepancy quasi-Monte Carlo achieves the more favorable $O((\log N)^d/N)$ order in low-dimensional settings.

\subsection{\texorpdfstring{Interpretation for PINN Search Efficiency}{Interpretation for PINN Search Efficiency}}

In collocation-based PINN training, the optimizer acts on the discrete residual loss $\widehat{L}_{\mathrm{PDE}}(\theta)$ rather than on the exact integral $L_{\mathrm{PDE}}(\theta)$. A lower-discrepancy point set reduces local over-sampling and under-sampling, thereby decreasing coverage holes in the spatio-temporal domain. In this sense, the discrete residual average more stably represents the global residual integral, which can improve the efficiency of the collocation search performed by the optimizer.

The analysis above explains the search-efficiency advantage of Sobol collocation from the perspective of residual-integral approximation. In shock-dominated problems, the magnitude of this advantage can vary with the residual structure and benchmark difficulty, which is consistent with the numerical observations in Section~3.6: the improvement is more pronounced on Sod and more modest on Shu--Osher. Importantly, both the mathematical argument and the ablation results support the use of Sobol sampling as a principled and practically effective collocation strategy in the present UM-PINN framework.

\FloatBarrier
\section{\texorpdfstring{Implementation Details of the Additional Baseline Comparisons}{Implementation Details of the Additional Baseline Comparisons}}
\label{sec:appendix_additional_baselines}

This appendix documents the implementation protocols for the additional baseline comparisons introduced in this study. It specifies the comparator families implemented in the present code base, the benchmark-matched training budgets, and the fairness principles adopted to ensure that the empirical comparisons directly test the targeted methodological differences.

\subsection{\texorpdfstring{Causal-loss baseline}{Causal-loss baseline}}

The causal comparator in Section~3.8 is implemented as a causal-loss weighting baseline inspired by causality-respecting PINN training \cite{Wang2024CausalPINN}. In the present implementation, this comparator is evaluated on the same two 1D shock benchmarks as the full UM-PINN, under the same benchmark-specific training budgets. For Sod, both the full UM-PINN and the causal-loss baseline are trained for 15{,}000 epochs; for Shu--Osher, both are trained for 20{,}000 epochs.

This causal comparator augments the global-in-time residual training objective with an explicit causal weighting factor. In the Sod implementation, the per-time-slab residual losses are accumulated in temporal order and reweighted by an exponential factor of the form $\exp(-\varepsilon_{\mathrm{c}}\cdot \text{cumulative loss})$, where the default setting uses $\varepsilon_{\mathrm{c}}=1.0$. In the Shu--Osher implementation, the pointwise residuals are first binned in time and then weighted by the same causal-loss principle. This design provides a direct test of whether causal residual weighting improves performance relative to the proposed dual-modulation strategy under matched global-in-time PINN training conditions.

\subsection{\texorpdfstring{Residual-based adaptive refinement and shock-aware adaptive sampling}{Residual-based adaptive refinement and shock-aware adaptive sampling}}

The RAR comparator is implemented as a residual-driven adaptive collocation update motivated by residual-based adaptive sampling studies \cite{Wu2023AdaptiveSampling}. In the benchmark configuration for the 1D benchmarks, candidate collocation points are generated periodically, their PDE residuals are evaluated, and the next collocation set is updated by selecting high-score candidates. The implemented 1D RAR score is simply the residual magnitude, i.e., $\mathrm{score}=|R|$.

The shock-aware comparator extends this adaptive refinement logic by combining residual information with a shock-sensitive density-gradient indicator. For the 1D runs, the implemented score takes the form

{
\begin{equation}
\mathrm{score}_{\mathrm{shock}} = |R| + \lambda_{\mathrm{shock}}\,|\partial_x \rho|,
\end{equation}
}

with $\lambda_{\mathrm{shock}}=1.0$ in the recorded 1D configurations. The 1D RAR and shock-aware runs use the same candidate-pool size and update cadence in the implementation: 4{,}096 candidate points are evaluated, 256 points are selected per update, and the replacement step is triggered every 1{,}000 epochs starting from epoch 1{,}000. These settings establish a controlled adaptive-sampling comparison in which the residual-driven and shock-aware mechanisms are evaluated under the same candidate-pool size, update cadence, and training scaffold.

\subsection{\texorpdfstring{XPINN baseline}{XPINN baseline}}

The XPINN baseline is included as a representative domain-decomposition PINN comparator following Ref.~\cite{jagtap2020xpinns}. It is evaluated on the same 1D Sod and Shu--Osher benchmarks as the other additional comparators, and its quantitative and qualitative results are reported in Section~3.9.

\subsection{\texorpdfstring{Weak-form-like / conservative-inspired baseline}{Weak-form-like / conservative-inspired baseline}}

The weak-form-like comparator used in this study adopts a conservative-inspired control-volume construction motivated by recent conservative and variational PINN developments \cite{Jagtap2020cPINN,Kharazmi2021hpVPINN}. It augments the pointwise residual and initial-condition losses with a local control-volume conservation penalty, yielding the runner-level loss structure

{
\begin{equation}
L_{\mathrm{total}} = L_{\mathrm{pointwise\_pde}} + L_{\mathrm{ic}} + w_{\mathrm{cv}} L_{\mathrm{control\_volume}} .
\end{equation}
}

For the 1D runs, the control-volume term is evaluated by sampling local space--time boxes, averaging the conservative variables on the top and bottom faces, averaging the fluxes on the left and right faces, and penalizing the resulting conservation defect. The recorded 1D configurations use $w_{\mathrm{cv}}=1.0$, 16 control volumes per iteration, and 32 quadrature-style sample points per control volume. This construction defines the weak-form-like / conservative-inspired comparator used in the present study.

\FloatBarrier
\clearpage

\bibliographystyle{elsarticle-num}
\bibliography{references}

\end{document}